\documentclass[12pt]{article}
\usepackage{amsmath,amssymb,bm,epsfig}

\renewcommand{\thefootnote}{\fnsymbol{footnote}}
\def\slash#1{\not\!\!#1}

\begin{document}

\title{
\begin{flushright}
\begin{minipage}{0.2\linewidth}
\normalsize
WUF-HEP-15-04 \\
EPHOU-15-006 \\*[50pt]
\end{minipage}
\end{flushright}
{\Large \bf 
Realistic three-generation models\\ 
from $SO(32)$ heterotic string theory
\\*[20pt]}}

\author{Hiroyuki~Abe$^{1,}$\footnote{
E-mail address: abe@waseda.jp},\ \ 
Tatsuo~Kobayashi$^{2}$\footnote{
E-mail address:  kobayashi@particle.sci.hokudai.ac.jp}, \ \ 
Hajime~Otsuka$^{1,}$\footnote{
E-mail address: hajime.13.gologo@akane.waseda.jp
}, \ and \
Yasufumi~Takano$^{2}$\footnote{
E-mail address: takano@particle.sci.hokudai.ac.jp}
\\*[20pt]
$^1${\it \normalsize 
Department of Physics, Waseda University, 
Tokyo 169-8555, Japan} \\
$^2${\it \normalsize 
Department of Physics, Hokkaido University, Sapporo 060-0810, Japan} \\*[50pt]}

\date{
\centerline{\small \bf Abstract}
\begin{minipage}{0.9\linewidth}
\medskip 
\medskip 
\small
We search for realistic supersymmetric 
standard-like models from $SO(32)$ heterotic string theory on factorizable tori with multiple magnetic fluxes. 
Three chiral ganerations of quarks and leptons are derived from the adjoint and 
vector representations of $SO(12)$ gauge groups 
embedded in $SO(32)$ adjoint representation.
Massless spectra of our models also include Higgs fields, which have 
desired Yukawa couplings to quarks and leptons at the tree-level. 
\end{minipage}
}

\begin{titlepage}
\maketitle
\thispagestyle{empty}
\clearpage
\tableofcontents
\thispagestyle{empty}
\end{titlepage}

\renewcommand{\thefootnote}{\arabic{footnote}}
\setcounter{footnote}{0}
\vspace{35pt}

\section{Introduction} 
Superstring theory is a good candidate for the unified 
theory of the gauge and gravitational interactions, and quark, lepton and Higgs fields. 
Indeed, 
there have been several approaches to derive the realistic string 
vacua by comparing the theoretical predictions with the 
data of the cosmological observations as well as the 
collider experiments which are known as the subjects of string cosmology 
and phenomenology. 

Beginning with the work of Ref.~\cite{Witten:1984dg}, there 
are much progresses to find the standard-like models from 
the $E_8 \times E_8$ heterotic string theory instead of the 
$SO(32)$ heterotic string theory. (See for a review, e.g. \cite{Ibanez:2012zz}.)
 This is because $E_8$ 
gauge group involves several candidates of the grand 
unified groups such as 
$E_6$, $SO(10)$ and $SU(5)$ as the subgroups of $E_8$ and the $E_8$ adjoint 
representation includes matter representations such as 
${\bf 27}$ of $E_6$, ${\bf 16}$ of $SO(10)$ and ${\bf 10}$ and $\bar {\bf 5}$ of $SU(5)$. 
However, in $SO(32)$ heterotic string theory, for 
example, the ${\bf 16}$ spinor representation of $SO(10)$ is not involved in 
the adjoint representation of $SO(32)$.
(In the framework of toroidal $Z_N$ orbifold, there 
are some possibilities to obtain the spinor representation of 
$SO$ groups as discussed in Ref.~\cite{Nilles:2006np}.) 
Therefore, as one of the procedures to find the realistic 
string vacua, we try to derive the (non-)supersymmetric 
standard-like models from the $SO(32)$ heterotic 
string theory without going through the 
grand unified groups. This approach might be 
useful to search for the realistic standard model, 
because the standard-like model given through 
the decomposition of GUT groups have the extra 
matters which should be decoupled from the 
low-energy dynamics in terms of some non-trivial mechanisms.

The standard model is a chiral theory.
Thus, the key point to realize the standard model is 
how to realize a chiral theory.
Toroidal compactification is simple, but it can not realize 
a chiral theory unless introducing additional backgrounds.
Orbifold and Calabi-Yau compactifications can lead to a chiral theory.
Toroidal compactification with magnetic fluxes can also lead to a chiral theory.
Here, we study such a background.
That is, our key ingredients are the multiple $U(1)$ magnetic fluxes 
inserted into $SO(32)$ gauge group. 
These magnetic fluxes 
are first discussed in Ref.~\cite{Witten:1984dg}, 
where the $SU(5)$ grand unified groups can be 
realized from the $SO(32)$ heterotic string theory. 
Furthermore, there are much progresses on the resolved toroidal orbifold in~\cite{Nibbelink:2012de} and on more general Calabi-Yau manifolds for $E_8 \times E_8$ and/or $SO(32)$ 
heterotic string theory via the spectral cover construction~\cite{Friedman:1997ih} and the 
extension of it~\cite{Blumenhagen:2005zg} ( see e.g., 
Refs.~\cite{Donagi:2000zf}).\footnote{Also the low-energy massless spectra were studied 
within the ten-dimensional $E_8 \times E_8$ theory on torus with magnetic fluxes from the field-theoretical viewpoint~\cite{Choi:2009pv}.}

In this paper, we study $SO(32)$ heterotic string theory on six-dimensional (6D) torus with magnetic fluxes, 
which is one of the simplest compactifications leading to a chiral theory.
Then, we search the models, where the unbroken gauge group includes 
$SU(3) \times SU(2) \times U(1)_Y$ and massless spectra correspond to 
three chiral generations of quarks and leptons.

The paper is organized as follows. In Sec.~\ref{sec:2}, we 
show our set-up and typical theoretical constraints which are required 
from the consistency of heterotic string theory. 
For example, 
in the  
standard embedding scenario of the Calabi-Yau compacfitication, 
 the internal gauge backgrounds are set to be equal to 
spin connections of the Calabi-Yau manifolds. 
On the other hand, in the  
non-standard embedding scenario, the gauge fields are 
not always identified as the spin connections of the internal 
manifold due to the existence of the fluxes. 
We discuss the consistency conditions for such fluxes on 6D torus
in Sec.~\ref{subsec:notation}. 
In addition, the $U(1)_Y$ gauge boson should be massless, even 
 if the consistent fluxes are inserted into $SO(32)$ gauge groups in order to derive the standard-like model gauge groups. 
Generically, $U(1)$ gauge bosons appeared in the low-energy 
effective theory couple to the universal and K\"ahler 
axions through the 
ten-dimensional (10D) Green-Schwarz term~\cite{Ibanez:1986xy} which implies that the linear combination of $U(1)$ gauge bosons may absorb these axions by their Stueckelberg couplings and become 
massive. Thus the axionic couplings of $U(1)_Y$ gauge boson 
should be absent, otherwise $U(1)_Y$ gauge boson would 
become massive as discussed in Sec.~\ref{subsec:GGS}. 
In addition to the merits of gauge symmetry breaking, the fluxes are 
important tools to realize the degenerate zero-modes, i.e.,  three generations of the elementary particles. 
In fact, in Sec.~\ref{subsec:4d}, the chiral theory with degenerate zero-modes can be obtained from the 
considerations of zero-mode wavefunctions on tori. 
At the same time, the existence of four-dimensional ($4$D) ${\cal N}=1$ supersymmetry (SUSY) depends on the ansatz of $U(1)$ fluxes due to the flux-induced Fayet-Iliopoulos terms. 

In Sec.~\ref{subsec:matter}, we discuss the concrete  embeddings of the standard model gauge groups into 
$SO(32)$ gauge group in terms of the multiple $U(1)$ 
fluxes.  The correct matter contents of the standard 
model are then derived from the adjoint and vector  representations of $SO(12)$ given by the subgroup of 
$SO(32)$. Since the number of generations corresponds 
to the number of $U(1)$ fluxes, we search for the 
desired matter contents of the standard model 
satisfying the $U(1)_Y$ massless conditions as well as 
the SUSY conditions as can be seen in Sec.~\ref{subsec:three}. 
In Sec.~\ref{subsec:threeK}, we further constrain the models by imposing 
the so-called K theory constraints. 
Finally, Sec.~\ref{sec:con} is devoted to the conclusion.
The normalization of $SO(32)$  generators and 
useful trace identities of $SO(32)$ gauge group are 
summarized in Appendices~\ref{app:normalization} and~\ref{app:trace}, 
respectively. 

\section{$SO(32)$ heterotic string theory on tori with $U(1)$ magnetic fluxes} 
\label{sec:2}
\subsection{Low-energy description of $SO(32)$ heterotic string theory} 
\label{subsec:notation}
We briefly review the $SO(32)$ heterotic string theory on a general 
complex manifold with multiple $U(1)$ magnetic fluxes. 
The notation is based on Refs.~\cite{Polchinsky,Blumenhagen:2005ga,Weigand:2006yj}. 
The low-energy effective action of $SO(32)$ heterotic string theory is given 
by 
\begin{align}
S_{\rm bos}&=\frac{1}{2\kappa_{10}^2}\int_{M^{(10)}} 
e^{-2\phi_{10}} \left[ R+4d\phi_{10} \wedge  \ast 
d\phi_{10} -\frac{1}{2}H\wedge \ast H \right] 
\nonumber\\
&-\frac{1}{2g_{10}^2} \int_{M^{(10)}} e^{-2\phi_{10}} 
{\rm tr}(F\wedge \ast F),
\label{eq:heterob}
\end{align}
which is the bosonic part of the action at 
the string frame in the notation of~\cite{Polchinsky}. 
The gravitational and Yang-Mills couplings are 
set by $2\kappa_{10}^2 =(2\pi)^7 (\alpha^{'})^4$ and 
$g_{10}^2 =2(2\pi)^7 (\alpha^{'})^3$ and 
$\phi_{10}$ denotes the ten-dimensional 
dilaton. Here the field-strength of $SO(32)$ 
gauge groups $F$ has the index of vector-representation. 
In what follows, ``tr" and ``Tr" represent for the trace in the vector and 
adjoint representation of the $SO(32)$ gauge group, respectively.  
In addition, $H$ denotes the heterotic three-form field strength defined by  
\begin{align}
H&=dB^{(2)} -\frac{\alpha^{'}}{4}(w_{\rm YM} -w_{L}),
\end{align}
where $w_{\rm YM}$ and $w_{L}$ are the gauge and 
gravitational Chern-Simons three-forms, respectively. 

From the action given by Eq.~(\ref{eq:heterob}), the kinetic term of the 
B-field is extracted as
\begin{align}
S_{\rm kin} +S_{\rm WZ}&= 
-\frac{1}{4\kappa_{10}^2}\int_{M^{(10)}} dB^{(2)}\wedge \ast dB^{(2)} 
-\sum_a N_a T_5 \int_{\Gamma_a} B^{(6)} \nonumber\\
&=-\frac{1}{4\kappa_{10}^2}\int_{M^{(10)}} dB^{(2)}\wedge \ast dB^{(2)} 
-\sum_a N_a T_5 \int_{M^{(10)}} B^{(6)} \wedge \delta(\Gamma_a), 
\label{eq:kinwz}
\end{align}
where we add the Wess-Zumino term which describes the magnetic 
sources for the Kalb-Ramond field $B^{(6)}$. Such sources correspond to the non-perturbative objects, i.e., 
the stacks of $N_a$ five-branes which wrap the holomorphic two-cycles 
$\Gamma_a$ and their tensions are given by $T_5=((2\pi)^5 (\alpha^{'})^3)^{-1}$. 
Here, $\delta (\Gamma_a)$ denote the Poinc\'are dual four-form 
of the two-cycles $\Gamma_a$. 

By employing the ten-dimensional Hodge duality, the Kalb-Ramond 
two-form $B^{(2)}$ and six-form $B^{(6)}$ are related as
\begin{align}
\ast dB^{(2)} =e^{2\phi_{10}} dB^{(6)},
\label{eq:Bhod}
\end{align}
and then the kinetic term of Kalb-Ramond field 
and Wess-Zumino term (\ref{eq:kinwz}) are rewritten as 
\begin{align}
S_{\rm kin} +S_{\rm WZ}&= -\frac{1}{4\kappa_{10}^2}\int_{M^{(10)}} 
e^{2\phi_{10}} dB^{(6)} \wedge \ast dB^{(6)} \nonumber\\ 
&+\frac{\alpha~{'}}{8\kappa_{10}^2}\int_{M^{(10)}}
B^{(6)} \wedge \left( {\rm tr}F^2 -{\rm tr}R^2 -4(2\pi)^2 \sum_a N_a \delta(\Gamma_a) \right),
\label{eq:kinwzd}
\end{align}
where 
$N_a =\pm1$ represent for the contributions of  
heterotic and anti-heterotic five-brane, respectively. 
The equation of motion of $B^{(6)}$ leads to the 
following tadpole condition of the NS-NS fluxes in 
the presence of five-branes, 
\begin{align}
d(e^{2\phi_{10}}\ast dB^{(6)}) =-\frac{\alpha^{'}}{4} 
\left({\rm tr}\bar{F}^2 -{\rm tr}{\bar R}^2 -
4(2\pi)^2 \sum_a N_a \delta(\Gamma_a) \right) 
=0,
\label{eq:tad}
\end{align}
in cohomology and where $\bar{F}$ stand for the 
gauge field strengths of the internal gauge 
fields whose gauge groups are embedded in 
$SO(32)$. 
When the extra-dimension is compactified on 
the flat space such as three 2-tori, 
$(T^2)_1\times (T^2)_2\times (T^2)_3$, the 
tadpole cancellation requires the following 
consistency conditions, 
\begin{align}
\int_{(T^2)_i\times (T^2)_j}
\left({\rm tr}\bar{F}^2 -4(2\pi)^2 \sum_a N_a 
\delta(\Gamma_a) \right)
=0,
\label{eq:tad2}
\end{align}
which should be satisfied on $(T^2)_i\times (T^2)_j$ 
with $i\neq j$, $i,j=1,2,3$. Thus  if the nonvanishing fluxes are not canceled by themselves, 
the non-perturbative objects would contribute to 
the cancellation of anomalies. It suggests that 
the modular invariance of heterotic string theory 
is recovered by the existence of these non-perturbative objects~\cite{Witten:1995gx,Duff:1996rs} which can 
be also realized in the framework of heterotic orbifold~\cite{Aldazabal:1997wi}.~\footnote{Even 
if the consistency condition is satisfied at 
the non-perturbative level, we have to care 
about the anomaly on heterotic 
five-branes and 
the global Witten anomaly is 
absent if the number of chiral fermions on 
the heterotic five branes is 
even~\cite{Witten:1982fp,Witten:1985xe}.}

\subsection{Generalized Green-Schwarz mechanism}
\label{subsec:GGS}
In addition to the consistency condition as discussed 
in Sec.~\ref{subsec:notation}, it must be ensured that 
our models do not have gauge 
and gravitational anomalies. 
In heterotic string theory, it is known that some gauge 
and gravitational anomalies are canceled by considering 
the following one-loop Green-Schwarz term 
at the string frame~\cite{Ibanez:1986xy}, 
\begin{align}
S_{\rm GS}=\frac{1}{24(2\pi)^5\alpha'}\int B^{(2)}\wedge X_8,
\label{eq:GS}
\end{align}
whose normalization factor is determined by the S-dual type 
I theory as shown in Appendix of \cite{Blumenhagen:2006ux} 
and the eight-form $X_8$ reads,
\begin{align}
X_8=\frac{1}{24}{\rm Tr}F^4-\frac{1}{7200}({\rm Tr}F^2)^2-
\frac{1}{240}({\rm Tr}F^2)({\rm tr}R^2) +\frac{1}{8}{\rm tr}R^4 
+\frac{1}{32}({\rm tr}R^2)^2.
\end{align}

Although the gauge and gravitational anomalies for the 
non-Abelian gauge groups are canceled 
by the above Green-Schwarz term~(\ref{eq:GS}) and the tadpole 
condition~(\ref{eq:tad}) as shown in Ref.~\cite{Witten:1984dg}, 
the anomalies relevant to the multiple Abelian gauge groups, which appear
in low-energy effective theory,  can be also canceled by same 
Green-Schwarz mechanism, for more details see 
Refs.~\cite{Blumenhagen:2005ga}. 
In fact, since we derive just the three-generation 
standard-like models, our phenomenological models do 
not receive these anomalies. 
However, as pointed out in Refs.~\cite{Blumenhagen:2005ga}, 
even if the Abelian gauge symmetries are anomaly-free, 
the Abelian gauge bosons may become massive due to the 
Green-Schwarz coupling given by Eq.~(\ref{eq:GS}). 
In order to ensure that the hypercharge 
gauge boson is massless, they should 
not couple to the axions which is hodge dual to the 
Kalb-Ramond field. 

For completeness, we define the hypercharge gauge group as the 
subgroup of $SO(32)$ as follows. 
The decomposition of the $SO(32)$ 
gauge group can be realized by inserting the multiple $U(1)$ 
constant magnetic fluxes as those satisfying 
\begin{align}
SO(32) \rightarrow 
SU(3)_C \otimes SU(2)_L \otimes_{a=1}^{13} U(1)_a.
\end{align}
Totally, $SO(32)$ has 16 Cartan elements, $H_i$ ($i=1,\cdots,16$).
We take the Cartan elements of $SU(3)$ along 
$H_1 -H_2$, $H_1 +H_2 -2H_3$ and Cartan element of $SU(2)$ as 
$H_5 - H_6$.
The other Cartan directions of $SO(32)$ are 
chosen as, 
\begin{align}
&U(1)_1:\,\,(0,0,0,0,1,1;0,0,\cdots, 0), 
\nonumber\\ 
&U(1)_2:\,\,(1,1,1,1,0,0;0,0,\cdots, 0), 
\nonumber\\
&U(1)_3:\,\,(1,1,1,-3,0,0;0,0,\cdots, 0),
\nonumber\\
&U(1)_4:\,\,(0,0,0,0,0,0;1,0,\cdots, 0),
\nonumber\\
&U(1)_5:\,\,(0,0,0,0,0,0;0,1,\cdots, 0),
\nonumber\\
& \qquad \vdots 
\nonumber\\
&U(1)_{13}:\,\,(0,0,0,0,0,0;0,0,\cdots, 1),
\label{eq:cartan}
\end{align}
in the basis $H_i$.
Then, we use the basis that non-zero roots have charge 
\begin{eqnarray}
(\underline{\pm 1, \pm 1, 0, \cdots, 0}),
\end{eqnarray}
under $H_i$ ($i=1,\cdots,16$), where the underline means any possible permutations.
The normalization of the Abelian gauge groups are discussed 
in the Appendix~\ref{app:normalization} and the concrete 
identification of standard model gauge groups and its representations 
are shown in Sec.~\ref{sec:embedding}. 
Note that some gauge groups would be enhanced to the larger 
one if any of $U(1)$ fluxes are absent or degenerate. 

When the $U(1)$ fluxes are inserted along the Cartan 
direction of $SO(32)$, the 
field strengths of $U(1)$s, $f$ are decomposed into the four-dimensional 
parts $f$ and extra-dimensional parts $\bar{f}$, 
\begin{align}
f \rightarrow f+\bar{f},
\end{align}
and then we can dimensionally reduce the one-loop 
Green-Schwarz term (\ref{eq:GS}) to 
\begin{align}
S_{\rm GS} = &\frac{1}{(2\pi )^3 l_s^2}
\int_{M^{(10)}} B^{(2)} \wedge \frac{1}{144}({\rm Tr}F\bar{f}^3)
\label{eq:GSA}\\
&-\frac{1}{(2\pi )^3 l_s^2}
\int_{M^{(10)}} B^{(2)} \wedge \frac{1}{2880}({\rm Tr}F\bar{f}) \wedge 
\left( \frac{1}{15}{\rm Tr}\bar{f}^2 +{\rm tr}{\bar R}^2\right) 
\label{eq:GSB}\\
&+\frac{1}{(2\pi )^3 l_s^2}
\int_{M^{(10)}} B^{(2)} \wedge 
\Bigl[\frac{1}{96}({\rm Tr}F^2\bar{f}^2) 
-\frac{1}{43200}({\rm Tr}F\bar{f})^2\Bigl] 
\label{eq:GSC}\\
&-\frac{1}{(2\pi )^3 l_s^2}
\int_{M^{(10)}} B^{(2)} \wedge 
\frac{1}{5760}({\rm Tr}F^2) \wedge 
\left( \frac{1}{15}{\rm Tr}\bar{f}^2 +{\rm tr}{\bar R}^2\right) 
\label{eq:GSD}\\
&+\frac{1}{(2\pi )^3 l_s^2}
\int_{M^{(10)}} B^{(2)} \wedge 
\frac{1}{384}({\rm tr}R^2) \wedge 
\left( {\rm tr}\bar{R}^2 -\frac{1}{15}{\rm Tr}{\bar f}^2\right) 
\label{eq:GSE}
\end{align}
where $l_s=2\pi \sqrt{\alpha^{\prime}}$, $F$ denote the 
field strengths of $SU(3)_C$, $SU(2)_L$, $U(1)_Y$. The explicit forms of traces appeared in Eqs.~(\ref{eq:GSA})-(\ref{eq:GSE}) 
are shown in Appendix~\ref{app:trace}. 

Before evaluating the mass term of $U(1)$ gauge bosons, 
for completeness, we show the definition of three 2-tori 
$(T^2)_i \simeq {\bf C}/\Lambda_i$ with $i=1,2,3$, 
where the lattices $\Lambda_i$ are generated by 
two vectors $e_i=2\pi R_i$ and $e_i=2\pi R_i\tau_i$. 
Here, $R_i$ and $\tau_i$ are the radii and complex structure 
moduli of $(T^2)_i$, respectively. 
The metrics of three 2-tori are then given by 
\begin{align}
&ds_6^2=g_{mn}dx^{m}dx^{n} =2h_{i\bar{j}}dz^idz^{\bar{j}}, 
\\
&g_{mn}=
\begin{pmatrix}
g^{(1)} & 0 & 0\\
0 & g^{(2)} & 0 \\
0 & 0 & g^{(3)}
\end{pmatrix}
,\,\,\,
h_{i\bar{j}}=
\begin{pmatrix}
h^{(1)} & 0 & 0\\
0 & h^{(2)} & 0 \\
0 & 0 & h^{(3)}
\end{pmatrix}
,
\end{align}
where $x^{m}$ are the coordinates of $T^2$ with 
$m,n=4,5,6,7,8,9$, $z^i=x^{2+2i}+\tau^{i} x^{3+2i}$ 
and the rank $2$ diagonal matrices $g^{(i)}$ and $h^{(i)}$ 
are given by 
\begin{align}
g^{(i)}=(2\pi R_i)^2 
\begin{pmatrix}
1 & {\rm Re}\,\tau_i \\
{\rm Re}\,\tau_i & |\tau_i|^2 
\end{pmatrix}
,\,\,\,
h^{(i)}=(2\pi R_i)^2
\begin{pmatrix}
0 & 1/2 \\
1/2 & 0 
\end{pmatrix}
.
\label{eq:metric}
\end{align}

From this expression, we expand the Kalb-Ramond field $B^{(2)}$ 
and internal $U(1)_a$ field strengths $\bar{f}_a$, $(a=1,\cdots, 13)$ 
in the basis of K\"ahler forms, 
$w_i =idz^i\wedge d\bar{z}^i/(2\,{\rm Im}\tau^{(i)})$ 
on tori $(T^2)_i$ derived from the metrics (\ref{eq:metric}), 
\begin{align}
&B^{(2)} =b_S^{(2)} +l_s^2\sum_{i=1}^{3} b_i^{(0)} w_i, 
\nonumber\\
&\bar{f}_a=2\pi \sum_{i=1}^{3} m_a^{(i)} w_i,
\label{eq:expansion}
\end{align}
where $m_a^{(i)}$ are the integers or half-integers  determined by Dirac quantization condition. 
Since Dirac quantization is satisfied in the 
adjoint representation of $SO(32)$, 
the factional numbers of $m_a^{(i)}$ can be allowed 
as pointed out in Ref.~\cite{Witten:1984dg}. 
From the Eqs.~(\ref{eq:GSA}) and (\ref{eq:GSB}), we can 
extract the Stueckelberg couplings,
\begin{align}
\frac{1}{3(2\pi )^3 l_s^2} 
\int b_S^{(2)} \wedge &\Bigl[ 
{\rm tr}T_1^4{\bar f}_1^3f_1 +
\left({\rm tr}T_2^4{\bar f}_2^3 +
3({\rm tr}T_2^2T_3^2){\bar f}_2{\bar f}_3^2 +
({\rm tr}T_2T_3^3){\bar f}_3^3 
\right) f_2 
\nonumber\\
&+\left(
{\rm tr}T_3^4{\bar f}_3^3 +
3({\rm tr}T_2T_3^3){\bar f}_2{\bar f}_3^2 +
3({\rm tr}T_2^2T_3^2){\bar f}_2^2{\bar f}_3
\right) f_3 
+\sum_{c=4}^{13}{\rm tr}T_c^4{\bar f}_c^3f_c\Bigl],
\end{align}
where the trace identities are employed as shown in 
Appendix~\ref{app:trace}. 
If the $U(1)$ gauge fields couple to the universal axion 
$b_S^{(0)}$ which is the hodge dual of the tensor field $b_S^{(2)}$, 
one of the multiple $U(1)$ gauge fields absorbs the 
universal axion and become massive. 
In our model, since the hypercharge $U(1)_Y$ is identified 
as the linear combinations of multiple $U(1)$s, i.e., 
$U(1)_Y=\frac{1}{6}(U(1)_3 +3\sum_{c}U(1)_c)$ 
as shall be discussed in Sec.~\ref{sec:embedding}~\footnote{In the 
definition of $U(1)_Y$, the summation over $c$ depends on the 
models as shown in Sec.~\ref{sec:embedding}}, 
the $U(1)_Y$ gauge field becomes massless under 
the condition
\begin{align}
&6{\rm tr}(T_3^4) m_3^{(1)}m_3^{(2)}m_3^{(3)} 
+3{\rm tr}(T_2T_3^3) d_{ijk}m_2^{(i)}m_3^{(j)}m_3^{(k)}
+3{\rm tr}(T_2^2T_3^2) d_{ijk}m_2^{(i)}m_2^{(j)}m_3^{(k)}
\nonumber\\
&+18\sum_{c} {\rm tr}(T_c^4)m_c^{(1)}m_c^{(2)}m_c^{(3)}=0,
\label{eq:massless1}
\end{align} 
which means no interaction between 
$U(1)_Y$ and the universal axion $b_S^{(0)}$. 
Here the following formulas are satisfied 
$\int_{T^2\times T^2\times T^2} 
{\bar f}_a^3=(2\pi)^3 d_{ijk}m_a^{(i)}m_a^{(j)}m_a^{(k)}
=6(2\pi)^3 m_a^{(1)}m_a^{(2)}m_a^{(3)}$ with 
the non-vanishing intersection numbers of 2-tori, $d_{ijk}=1$ 
($i\neq j\neq k$). 

Except for the universal axion, there are other axions associated with the internal cycles, that is, K\"ahler axions which couple to the $U(1)$ gauge 
bosons originated from the action given by 
Eq.~(\ref{eq:kinwzd}). 
Along with the Kalb-Ramond field $B^{(2)}$, we 
expand the dual field $B^{(6)}$ as 
\begin{align}
B^{(6)} =l_s^6b_0^{(0)}{\rm vol}_6 
+l_s^4\sum_{k=1}^{3} b_k^{(2)} \hat{w}_k, 
\end{align}
where $\hat{w}_k$ are the Hodge dual four-forms of the 
K\"ahler forms,
\begin{align}
&\hat{w}_k =\frac{d_{kij}}{2} i\frac{dz^i\wedge d\bar{z}^i}{2\,{\rm Im}\tau^{(i)}} \wedge i\frac{dz^j\wedge d\bar{z}^j}{2\,{\rm Im}\tau^{(j)}},
\end{align}
which are defined as those satisfying 
$\int_{T^2\times T^2 \times T^2} w_i 
\wedge \hat{w}_{j} =\delta_{ij}$. 
After inserting these expressions into the action 
given by Eq.~(\ref{eq:kinwzd}), 
we can extract the mass terms of the $U(1)$ gauge bosons,
\begin{align}
&\frac{1}{l_s^2} 
\int b_{i}^{(2)} \wedge \sum_{a=1}^{13}{\rm tr}(T_a^2)f_a m_{a}^{(i)}.
\end{align}
In the same way as the case of universal axion, 
the $U(1)_Y$ gauge field should not couple to the 
K\"ahler axions, otherwise it becomes massive. 
Thus the $U(1)_Y$ gauge boson is massless under  
the following condition,
\begin{align}
&{\rm tr}(T_3^2)m_3^{(i)} +3\sum_{c=4}^{13} {\rm tr}(T_c^2)m_c^{(i)}=0,
\label{eq:massless2}
\end{align} 
with $i=1,2,3$. 

As a step to realize the realistic models, the massless 
conditions for $U(1)_Y$ gauge boson given by Eqs.~(\ref{eq:massless1}) 
and (\ref{eq:massless2}) 
should be satisfied. It is remarkable that these $U(1)$ fluxes 
are sensitive to the consistency 
condition given by Eq.~(\ref{eq:tad2}) 
as shown in the Sec.~\ref{subsec:notation}. 
When the heterotic five-branes are absent in our system, 
the following conditions, 
\begin{align}
&\sum_{a=1}^{13}{\rm tr}(T_a^2)m_a^{(i)}m_a^{(j)} =0,
\,\,i\neq j,\,\,\,(i,j=1,2,3),
\end{align}
are required from the consistencies of heterotic string theory,   otherwise the NS-NS tadpole 
could be canceled by the existence of heterotic 
five-branes.

\subsection{The chiral fermions and degenerate 
zero-modes}
\label{subsec:4d}
The heterotic string theory on three 2-tori has ${\cal N}=4$ 
supersymmetry in the language of $4$D supercharges 
which have to be broken to at least ${\cal N}=1$ 
supersymmetry in the four-dimension, otherwise 
the chiral matters do not appear in the low-energy 
effective theory. Although it is known that there are much progresses 
in the framework of toroidal orbifold, in this paper, 
we focus on the realization of chiral fermions 
by employing the multiple $U(1)$ fluxes as discussed in this section.~\footnote{Although the gauge sector still remains $4$D ${\cal N}=4$ SUSY, it could be 
broken to ${\cal N}=1$ SUSY by extending our 
system to the toroidal orbifold with trivial gauge 
embedding. The (anti-) heterotic five-branes would break 
(all) partial SUSY. It is then expected that the heterotic 
five branes compensate the moduli invariance even 
if the moduli invariance is violated at the string 
tree-level.} 

First we define the $10$D Majorana-Weyl spinor $\lambda$ which 
satisfies the Majorana-Weyl condition, 
\begin{align}
\Gamma \lambda =\lambda,
\end{align}
where $\Gamma$ is the $10$D chirality matrix. The following analysis 
is based on Ref.~\cite{Cremades:2004wa}. In order to discuss 
the $4$D chirality, we decompose the $10$D Majorana-Weyl spinor 
$\lambda$ into four $4$D Weyl spinors $\lambda_0$ 
and $\lambda_i$ with $i=1,2,3$ as the representation of 
$SU(4)\simeq SO(6)$. 
The $10$D chirality matrix $\Gamma$ is also decomposed into the product of three $2$D 
chirality operators , $\Gamma_i=-i\Gamma_i^1\Gamma_i^2$ on $(T^2)_i$, 
where 
\begin{align}
\Gamma_i^1 =
\begin{pmatrix}
0 & 1 \\
1 & 0
\end{pmatrix}
,\,\,\,
\Gamma_i^2 =
\begin{pmatrix}
0 & -i \\
i & 0
\end{pmatrix}
,
\label{eq:gammaflat}
\end{align}
satisfying the Clifford algebra. 
Then the $4$D chirality is fixed as 
\begin{equation}
\Gamma^i \lambda_0 =\lambda_0,
\,\,\,
\Gamma^i \lambda_j =\left\{ 
\begin{array}{l}
+\lambda_j\,\,\, (i=j), \\
-\lambda_j\,\,\, (i\neq j),
\end{array}
\right.
\end{equation}
which lead to the following $4$D Weyl spinors,
\begin{equation}
\lambda_0 =\lambda_{+++}, \,\,\,
\lambda_1 =\lambda_{+--}, \,\,\,
\lambda_2 =\lambda_{-+-}, \,\,\,
\lambda_3 =\lambda_{--+},
\label{eq:chirality}
\end{equation}
where the subscript indexes denote the eigenvalues 
of $\Gamma^i$ with $i=1,2,3$. 
When we insert the magnetic fluxes on three 2-tori, 
one of the four $4$D Weyl spinors would be chosen.
In order to prove the above statements, 
we show the zero-mode wavefunction 
of fermions originating from the $10$D gaugino field 
by solving their Dirac equations.

The zero-modes of $10$D gaugino field $\lambda$ and 
gauge field $A_M$ are defined through the following 
decompositions,
\begin{align}
&\lambda (x^\mu, z^i) =\sum_n \chi_n (x^\mu) 
\otimes \psi_n^{(1)} (z^1) \otimes \psi_n^{(2)} (z^2) 
\otimes \psi_n^{(3)} (z^3),
\nonumber\\
&A_M (x^\mu, z^i) =\sum_n \varphi_{n,M} (x^\mu ) 
\otimes \phi_{n,M}^{(1)} (z^1) \otimes \phi_{n,M}^{(2)} (z^2) 
\otimes \phi_{n,M}^{(3)} (z^3),
\end{align}
where $M=0,1,\cdots,9$ and $x^\mu$, $\mu=0,1,2,3$ are the coordinates of the $4$D spacetime. 
The zero-modes of gaugino fields, 
$\psi_0^{(i)}(z^i)$ are expressed as 
\begin{align}
\psi_0^{(i)}(z^i) =
\begin{pmatrix}
\psi_+^{(i)} (z^i)\\
\psi_-^{(i)} (z^i)
\end{pmatrix}
,
\end{align}
where hereafter we omit the subscript $0$ of 
the zero-modes, that is, $\psi^{(i)}(z^i)=\psi_0^{(i)}(z^i)$. 
On the other hand, the extra dimensional 
components of $U(1)_a$ gauge backgrounds $A_a^{(i)}(z^i)$ 
($a=1,2,\cdots, 13)$ are given by 
\begin{align}
A_a^{(i)}(z^i) =\frac{\pi m_a^{(i)}}{{\rm Im}\,\tau_i} 
{\rm Im}\,({\bar z}_i dz_i),
\label{eq:gaugefield}
\end{align}
which lead to the magnetic fluxes given by 
Eq.~(\ref{eq:expansion}) along the Cartan 
direction of $SO(32)$. Here and hereafter, we multiply the $U(1)_a$ magnetic fluxes $m_a^{(i)}$ by their corresponding normalization factors. 

Then the zero-mode equations of 
fermions $\psi^{(i)}(z^i)$ with the $U(1)_a$ charge $q_a$ 
are given by 
\begin{align}
\slash{D}_i \psi^{(i)}(z^i) =(\Gamma^{z^i} \nabla_{z^i} 
+\Gamma^{{\bar z}^i} \nabla_{{\bar z}^i}) \psi^{(i)}(z^i)=0
\end{align}
where the Gamma matrices and covariant derivatives 
in terms of the complex coordinates, ($z^i,{\bar z}^i$) 
are defined as
\begin{align}
\Gamma^{z^i} =\frac{1}{2\pi R_i}
\begin{pmatrix}
0 & 2 \\
0 & 0
\end{pmatrix}
,\,\,\,
\Gamma^{{\bar z}^i} =\frac{1}{2\pi R_i}
\begin{pmatrix}
0 & 0 \\
2 & 0
\end{pmatrix}
, 
\end{align}
which can be derived from the 
Gamma matrices in flat space~(\ref{eq:gammaflat}) 
and the metric of torus~(\ref{eq:metric}) 
and 
\begin{align}
&\nabla_{z^i} =\partial_{z^i} -iq_a (A_{a}^{(i)})_{z^i},
\nonumber\\
&\nabla_{{\bar z}^i} =\partial_{{\bar z}^i} 
-iq_a (A_{a}^{(i)})_{{\bar z}^i}. 
\end{align}
The spin connections are vanishing due to the 
topology of tori. 
Thus the Dirac equations on $(T^2)_i$ 
are rewritten as 
\begin{align}
&\left( {\bar \partial}_{{\bar z}^i} +
\frac{\pi q^am_a^i}{2{\rm Im}\,\tau_i}z^i \right)
\psi_+^{(i)}(z^i, {\bar z}^i)=0,
\nonumber\\
&\left( \partial_{z^i} -
\frac{\pi q^am_a^i}{2{\rm Im}\,\tau_i}{\bar z}^i \right)
\psi_-^{(i)}(z^i, {\bar z}^i)=0.
\label{eq:diraceq}
\end{align} 
Then $\psi_+^{(i)}(z^i, {\bar z}^i)$ has zero-modes only if  $M^i=q_am_a^i >0$, whereas 
$\psi_-^{(i)}(z^i, {\bar z}^i)$ has zero-modes only if $M^i<0$.
In both cases, the wavefunctions have $|M^i|$ 
independent solutions as the solution of Dirac 
equations~(\ref{eq:diraceq}). Hence the number of 
generations of zero-modes, $M$ is given by 
the product of $|M^i|$, 
that is, $M=|M^1||M^2||M^3|$. (This result is consistent with that of the index theorem.) 
Since the nonvanishing fluxes $|M^i|$ select 
one of the two chiralities on $(T^2)_i$, i.e., 
$\psi_+^{(i)}$ or $\psi_-^{(i)}$, non vanishing fluxes on 
three 2-tori lead to the chiral spectrum 
as can be seen in Eq.~(\ref{eq:chirality}). 

However, such magnetic fluxes may break 
all ${\cal N}=4$ SUSY through the D-terms or 
Fayet-Iliopoulos terms in the 
language of $4$D ${\cal N}=1$ SUSY. 
When ${\cal N}=1$ SUSY is preserved in 
the system, the vanishing D-terms imply
that the hermitian Yang-Mills equations for 
the $U(1)_a$ field strengths 
should be satisfied at the vacuum, 
\begin{align}
g^{i{\bar j}}({\bar f}_a)_{i{\bar j}}=0. 
\end{align}
In our set-up, 
these conditions are equal to 
\begin{align}
\sum_{i=1}^3\frac{m_a^i}{{\cal A}_i}=0,
\label{eq:SUSYcond}
\end{align}
where ${\cal A}_i=(2\pi R_i)^2{\rm Im}\,\tau_i$ 
are the areas of tori, $(T^2)_i$. 
Indeed, when these conditions are satisfied, massless 
scalar fields appear for $A_M$ $(M=4,\cdots,9)$, and they 
correspond to superpartners of the above massless fermions.
At the perturbative level, the D-term 
conditions receive at most one-loop 
corrections~\cite{Fischler:1981zk} which 
have the dilaton dependence. 

Finally, we comment on the Wilson lines 
which play a role of breaking the 
gauge group into its subgroups without 
changing the rank of gauge groups. 
In fact, when we introduce the Wilson lines 
$\zeta_a^{(i)}$, along the $U(1)_a$ directions, the internal 
components of $U(1)_a$ gauge fields take 
the following shifts compared to 
Eq.~(\ref{eq:gaugefield}), 
\begin{align}
A_a(z^i) =\frac{\pi m_a^{(i)}}{{\rm Im}\,\tau_i} 
{\rm Im}\,(({\bar z}_i +{\bar \zeta}_a^{(i)})dz_i),
\end{align}
which modify the zero-mode wavefunctions 
determined by the Dirac equations~(\ref{eq:diraceq}), 
whereas the number of zero-modes and $U(1)$ 
fluxes are not modified. 
When we evaluate the values of Yukawa couplings, 
such Wilson lines would give significant effects.

\section{Three-generation models in the $SO(32)$ heterotic string theory}
\label{sec:embedding}
\subsection{Matter content}
\label{subsec:matter}
In this section, we show the concrete decomposition 
of $SO(32)$ gauge group into the standard model 
gauge groups and then the parts of adjoint representation of 
$SO(32)$ are identified as the matter contents of the standard 
model. 
As the first step to obtain the standard model gauge 
groups, we consider the decomposition of $SO(32)$  illustrated as
\begin{align}
SO(32) &\rightarrow SO(12) \otimes SO(20), 
\nonumber\\
496 &\rightarrow (1, 190) \oplus (12_v, 20_v)
\oplus (66, 1),
\end{align}
where the multiple $U(1)$ fluxes are assumed along the 
Cartan directions of $SO(32)$. 

In order to derive the matter contents of the standard 
model, we examine whether the adjoint representation of 
$SO(12)$ involves the candidates of elementary particles or not. 
When we put three $U(1)_{1,2,3}$ fluxes along the 
Cartan directions of $SO(12)$ gauge group, it is found that 
$SO(12)$ involves the candidates of $SU(3)_C$ and $SU(2)_L$, 
\begin{align}
SO(12) 
&\rightarrow SO(8) \otimes SU(2)_L \otimes U(1)_1 
\rightarrow SU(4) \otimes U(1)_2 \otimes SU(2)_L \otimes U(1)_1 
\nonumber\\
&\rightarrow SU(3)_C \otimes U(1)_3 \otimes U(1)_2 
\otimes SU(2)_L \otimes U(1)_1,
\end{align}
where the Cartan directions of $U(1)_{1,2,3}$ are 
given by Eq.~(\ref{eq:cartan}) .
Then the adjoint representation of $SO(12)$
is decomposed as 
\begin{align}
66 \left\{
\begin{array}{l}
(28,1)_0 
\left\{
\begin{array}{l}
(15,1)_{0,0} 
\left\{
\begin{array}{l}
(8,1)_{0,0,0}\\
(3,1)_{0,0,4}\\
({\bar 3},1)_{0,0,-4}\\
(1,1)_{0,0,0}\\
\end{array}
\right.\\
(6,1)_{0,2}
\left\{
\begin{array}{l}
(3,1)_{0,2,-2}\\
({\bar 3},1)_{0,2,2}
\end{array}
\right.\\
({\bar 6},1)_{0,-2}
\left\{
\begin{array}{l}
(3,1)_{0,-2,-2}\\
({\bar 3},1)_{0,-2,2}
\end{array}
\right.\\
(1,1)_{0,0,0} 
\end{array}
\right.\\
(8_v, 2)_1 
\left\{
\begin{array}{l}
(4,2)_{1,1} 
\left\{
\begin{array}{l}
(3,2)_{1,1,1}\\
(1,2)_{1,1,-3}
\end{array}
\right.\\
({\bar 4},2)_{1,-1}
\left\{
\begin{array}{l}
({\bar 3},2)_{1,-1,-1}\\
(1,2)_{1,-1,3}
\end{array}
\right.\\
\end{array}
\right.\\
(8_v, 2)_{-1} 
\left\{
\begin{array}{l}
(4,2)_{-1,1} 
\left\{
\begin{array}{l}
(3,2)_{-1,1,1}\\
(1,2)_{-1,1,-3}
\end{array}
\right.\\
({\bar 4},2)_{-1,-1}
\left\{
\begin{array}{l}
({\bar 3},2)_{-1,-1,-1}\\
(1,2)_{-1,-1,3}
\end{array}
\right.\\
\end{array}
\right.\\
(1,3)_{0,0,0} \\
(1,1)_{2,0,0} \\
(1,1)_{-2,0,0} \\
(1,1)_{0,0,0} 
\end{array}
\right.
,
\end{align}
which are singlets of $SO(20)$, 
where the subscript indices denote the $U(1)_{1,2,3}$ charge $q_{1,2,3}$. The normalization of $U(1)$ generators 
are given by Appendix~\ref{app:normalization}. 
Thus when we identify the hypercharge as $U(1)_Y=U(1)_3/6$, 
we can extract the candidates of the quarks, charged leptons 
and/or Higgs,
\begin{equation}
\begin{array}{lll}
Q:\,\left\{
\begin{array}{l}
Q_1=(3,2)_{1,1,1}\\
Q_2=(3,2)_{-1,1,1}
\end{array}
\right.
,&
L:\,\left\{
\begin{array}{l}
L_1=(1,2)_{1,1,-3}\\
L_2=(1,2)_{-1,1,-3}
\end{array}
\right.
,&
u_R^c:\,\left\{
\begin{array}{l}
u_{R_1}^c=({\bar 3},1)_{0,0,-4}
\end{array}
\right.
,\\
d_R^c:\,\left\{
\begin{array}{l}
d_{R_1}^c=({\bar 3},1)_{0,2,2}\\
d_{R_2}^c=({\bar 3},1)_{0,-2,2}
\end{array}
\right.
,&
n_1=(1,1)_{2,0,0}.
& 
\end{array}
\end{equation}
As shown in the above analysis, the adjoint representation of $SO(12)$, $66$ does not involve the candidate of right-handed 
leptons. Therefore, we further decompose the $SO(20)$ 
gauge group into $U(1)_{4,5,\cdots,13}$ gauge groups, 
\begin{align}
SO(20) \rightarrow U(1)_4 \otimes \cdots \otimes U(1)_{13},
\end{align}
where the nonvanishing $U(1)$ fluxes along 
all $U(1)_{4,\cdots,13}$ directions are inserted 
shown in Eq.~(\ref{eq:cartan}). Now $SO(2)$ 
is identified as $U(1)$. 
The vector representation and 
the singlet of $SO(12)$, $12_v$ and $1$ give the 
suitable matter contents, i.e., right-handed quarks and 
leptons, charged-leptons and/or Higgs, 
\begin{align}
(12_v, 20_v) &\rightarrow 
\left\{ 
\begin{array}{l}
L_3^a =(1,2)_{1,0,0;\underline{-1,0,\cdots,0}}\\
L_4^a =(1,2)_{-1,0,0;\underline{-1,0,\cdots,0}}\\
u_{R_2}^{c\,\,a}=({\bar 3},1)_{0,-1,-1;\underline{-1,0,\cdots,0}}\\
d_{R_3}^{c\,\,a}=({\bar 3},1)_{0,-1,-1;\underline{1,0,\cdots,0}}\\
e_{R_1}^{c\,\,a}=(1,1)_{0,-1,3;\underline{1,0,\cdots,0}}\\
n_{2}^{c\,\,a}=(1,1)_{0,-1,3;\underline{-1,0,\cdots,0}}\\
\end{array}
\right.
,\,(a=4,5,\cdots, 13),
\nonumber\\
(1, 190) &\rightarrow 
\left\{ 
\begin{array}{l}
e_{R_2}^{c\,\,ab}=(1,1)_{0,0,0;\underline{1,1,0,\cdots,0}}\\
n_{3}^{c\,\,ab}=(1,1)_{0,0,0;\underline{1,-1,0,\cdots,0}}\\
\end{array}
\right.
,\,(a,b=4,5,\cdots, 13,\,a\neq b),
\end{align}
where the underlines for $U(1)_{4,5,\cdots,13}$ charge 
$q_{4,5,\cdots,13}$ denote all the possible permutations. 
It is remarkable that the correct $U(1)_Y$ 
charge can be also realized as
\begin{align}
U(1)_Y=\frac{1}{6} \left( U(1)_3 +3\sum_{c=4}^{13}U(1)_c\right).
\end{align}

\subsection{Three-generation models}
\label{subsec:three}
Since the matter contents of the standard model are 
correctly identified in the previous section, 
we show the number of generations for each 
representation in this section. 

As discussed in Sec.~\ref{subsec:4d}, the $U(1)$ fluxes generate 
the degenerate zero-modes if these zero-modes have 
$U(1)$ charges. It implies that the number of generations for 
the representations embedded in the adjoint and vector 
representations of 
$SO(12)$, $66$ and $12_v$ are determined by the following formulas, 
\begin{equation}
\begin{array}{ll}
m_{Q_1} =\prod_{i=1}^3 m_{Q_1}^i 
=\prod_{i=1}^3 (m_{1}^i+m_2^i+m_3^i), &
m_{Q_2} =\prod_{i=1}^3 m_{Q_2}^i 
=\prod_{i=1}^3 (-m_{1}^i+m_2^i+m_3^i),\\
m_{L_1} =\prod_{i=1}^3 m_{L_1}^i 
=\prod_{i=1}^3 (m_{1}^i+m_2^i-3m_3^i),&
m_{L_2} =\prod_{i=1}^3 m_{L_2}^i 
=\prod_{i=1}^3 (-m_{1}^i+m_2^i-3m_3^i),\\
m_{u_{R_1}^c} =\prod_{i=1}^3 m_{u_{R_1}^c}^i 
=\prod_{i=1}^3 (-4m_3^i),&
m_{n_{1}} =\prod_{i=1}^3 m_{n_{1}}^i 
=\prod_{i=1}^3 (2m_1^i),\\
m_{d_{R_1}^c} =\prod_{i=1}^3 m_{d_{R_1}^c}^i 
=\prod_{i=1}^3 (2m_2^i+2m_3^i),&
m_{d_{R_2}^c} =\prod_{i=1}^3 m_{d_{R_2}^c}^i 
=\prod_{i=1}^3 (-2m_2^i+2m_3^i),
\label{eq:generationadj}
\end{array}
\end{equation}
and 
\begin{equation}
\begin{array}{ll}
m_{L_3^a} =\prod_{i=1}^3 m_{L_3^a}^i 
=\prod_{i=1}^3 (m_{1}^i-m_a^i), & 
m_{L_4^a} =\prod_{i=1}^3 m_{L_4^a}^i 
=\prod_{i=1}^3 (-m_{1}^i-m_a^i),\\
m_{u_{R_2}^{c\,a}} =\prod_{i=1}^3 m_{u_{R_2}^{c\,a}}^i 
=\prod_{i=1}^3 (-m_2^i-m_3^i-m_a^i), & 
m_{d_{R_3}^{c\,a}} =\prod_{i=1}^3 m_{d_{R_3}^{c\,a}}^i 
=\prod_{i=1}^3 (-m_2^i-m_3^i+m_a^i),\\
m_{e_{R_1}^{c\,a}} =\prod_{i=1}^3 m_{e_{R_1}^{c\,a}}^i 
=\prod_{i=1}^3 (-m_2^i+3m_3^i+m_a^i), & 
m_{n_2^{a}} =\prod_{i=1}^3 m_{n_2^{a}}^i 
=\prod_{i=1}^3 (-m_2^i+3m_3^i-m_a^i),
\label{eq:generatonvec}
\end{array}
\end{equation}
respectively. 

Now we are ready to search for the realistic 
three-generation models in the framework of 
$SO(32)$ heterotic string theory. 
In the light of $U(1)_Y$ massless conditions 
given by Eqs.~(\ref{eq:massless1}) and (\ref{eq:massless2}), 
the nonvanishing $U(1)_3$ fluxes seem to violate these 
massless conditions. Therefore, in this paper, 
we restrict ourselves to the case 
that $U(1)_3$ fluxes are absent in our system, 
which lead to no chiral generations of right-handed 
quarks, $u_{R}^c$ and $d_R^c$ from the adjoint representation of $SO(12)$ as can be seen in 
Eq.~(\ref{eq:generationadj}). 
Only left-handed quarks $Q$ and charged-leptons $L$ 
are then generated from the adjoint representation of 
$SO(12)$. 
As for the left-handed quarks, $Q$, 
there are two possibilities to reproduce 
the three generations of $Q$,
\begin{equation}
{\rm Type\,A}:\,(m_{Q_1},m_{Q_2})=(2,1),\quad
{\rm Type\,B}:\,(m_{Q_1},m_{Q_2})=(3,0),
\label{eq:case}
\end{equation}
without loss of generality, because we can 
exchange $m_{Q_1}$ and $m_{Q_2}$ under flipping 
the sign of $m_{1}^i$ with $i=1,2,3$. In both 
cases, the possible $U(1)$ fluxes are summarized in 
 Tables~\ref{table:A} and \ref{table:B} and in the 
case of Type B, it 
is restricted within the range of $-2\leq m_{Q_2}^i \leq 2$, $i=1,2,3$, for simplicity. 
In both tables, possible permutations among the first, second and third 2-tori 
are omitted.
Also, when we flip signs of magnetic fluxes in two of three 2-tori, 
we obtain the same generation number.
For example the magnetic fluxes, $( m^1_1, m^2_1,m^3_1)=(-3/2,0,1)$ $( m^1_2,m^2_2,m^3_2 )=(-1/2,-1,0)$, 
are obtained by flipping the signs of magnetic fluxes in the first and second 2-tori from 
ones in Table~\ref{table:A} and they lead to the same generation numbers.
We omit such possibilities in both tables.
\begin{table}[htb]
\begin{center}
\begin{tabular}{|c|c|} \hline
 $ \left( m^1_1, m^2_1,m^3_1 \right) $ & $ \left( m^1_2,m^2_2,m^3_2 \right) $\\ \hline \hline
			$( \frac{1}{2}, 0, 0 )$ & $( \frac{3}{2}, 1, 1 )$ \\
			$( 1, 1, \frac{1}{2} )$ & $( 0, 0, \frac{3}{2} )$ \\
			$( \frac{3}{2}, 1, 0 )$ & $( \frac{1}{2}, 0, 1 )$ \\\hline 
\end{tabular}
 \caption{The possible magnetic 
fluxes in Type A.
Possible permutations among the three 2-tori are omitted.
Certain types of sign flipping are also omitted.}
\label{table:A}
\end{center}
\end{table}
\begin{table}[htb]
\begin{center}
\begin{tabular}{|c|c|} \hline
 $ \left( m^1_1, m^2_1,m^3_1 \right) $ & $ \left( m^1_2,m^2_2,m^3_2 \right) $ \\ \hline \hline
	$( \frac{1}{2}, \frac{1}{2}, -\frac{1}{2} )$ & $( \frac{5}{2}, \frac{1}{2}, \frac{3}{2} )$ \\
			$( \frac{1}{2}, \frac{1}{2}, 0 )$ & $( \frac{5}{2}, \frac{1}{2}, 1 )$ \\
			$( \frac{1}{2}, \frac{1}{2}, \frac{1}{2} )$ & $( \frac{5}{2}, \frac{1}{2}, \frac{1}{2} )$ \\
			$( 1, \frac{1}{2}, -\frac{1}{2} )$ & $( 2, \frac{1}{2}, \frac{3}{2} )$ \\
			$( 1, \frac{1}{2}, 0 )$ & $( 2, \frac{1}{2}, 1 )$ \\
			$( 1, \frac{1}{2}, \frac{1}{2} )$ & $( 0, \frac{5}{2}, \frac{1}{2} )$ \\
			$( 1, \frac{1}{2}, \frac{1}{2} )$ & $( 2, \frac{1}{2}, \frac{1}{2} )$ \\
			$( 1, 1, \frac{1}{2} )$ & $( 2, 0, \frac{1}{2} )$ \\
			$( \frac{3}{2}, 0, -\frac{1}{2} )$ & $( \frac{3}{2}, 1, \frac{3}{2} )$ \\
			$( \frac{3}{2}, 0, 0 )$ & $( \frac{3}{2}, 1, 1 )$ \\
			$( \frac{3}{2}, \frac{1}{2}, -\frac{1}{2} )$ & $( \frac{3}{2}, \frac{1}{2}, \frac{3}{2} )$ \\
			$( \frac{3}{2}, \frac{1}{2}, 0 )$ & $( \frac{3}{2}, \frac{1}{2}, 1 )$ \\
			$( \frac{3}{2}, \frac{1}{2}, \frac{1}{2} )$ & $( -\frac{1}{2}, \frac{5}{2}, \frac{1}{2} )$ \\
			$( \frac{3}{2}, \frac{1}{2}, \frac{1}{2} )$ & $( \frac{3}{2}, -\frac{3}{2}, -\frac{3}{2} )$ \\
			$( \frac{3}{2}, \frac{1}{2}, \frac{1}{2} )$ & $( \frac{3}{2}, \frac{1}{2}, \frac{1}{2} )$ \\
			$( \frac{3}{2}, 1, -\frac{1}{2} )$ & $( \frac{3}{2}, 0, \frac{3}{2} )$ \\
			$( \frac{3}{2}, 1, 0 )$ & $( \frac{3}{2}, 0, 1 )$ \\
			$( \frac{3}{2}, 1, \frac{1}{2} )$ & $( -\frac{1}{2}, 2, \frac{1}{2} )$ \\
			$( \frac{3}{2}, 1, \frac{1}{2} )$ & $( \frac{3}{2}, 0, \frac{1}{2} )$ \\
			$( \frac{3}{2}, 1, 1 )$ & $( \frac{3}{2}, 0, 0 )$ \\
			$( \frac{3}{2}, \frac{3}{2}, -\frac{1}{2} )$ & $( \frac{3}{2}, -\frac{1}{2}, \frac{3}{2} )$ \\
			$( \frac{3}{2}, \frac{3}{2}, 0 )$ & $( \frac{3}{2}, -\frac{1}{2}, 1 )$ \\
			$( \frac{3}{2}, \frac{3}{2}, \frac{1}{2} )$ & $( \frac{3}{2}, -\frac{1}{2}, \frac{1}{2} )$ \\
			$( \frac{3}{2}, \frac{3}{2}, 1 )$ & $( \frac{3}{2}, -\frac{1}{2}, 0 )$ \\
			$( \frac{3}{2}, \frac{3}{2}, \frac{3}{2} )$ & $( \frac{3}{2}, -\frac{1}{2}, -\frac{1}{2} )$ \\
			$( 2, \frac{1}{2}, -\frac{1}{2} )$ & $( 1, \frac{1}{2}, \frac{3}{2} )$ \\
			$( 2, \frac{1}{2}, 0 )$ & $( 1, \frac{1}{2}, 1 )$ \\
			$( 2, \frac{1}{2}, \frac{1}{2} )$ & $( 1, \frac{1}{2}, \frac{1}{2} )$ \\
			$( 2, 1, \frac{1}{2} )$ & $( 1, 0, \frac{1}{2} )$ \\
			$( 2, \frac{3}{2}, \frac{1}{2} )$ & $( 1, -\frac{1}{2}, \frac{1}{2} )$ \\
			$( \frac{5}{2}, \frac{1}{2}, -\frac{1}{2} )$ & $( \frac{1}{2}, \frac{1}{2}, \frac{3}{2} )$ \\
			$( \frac{5}{2}, \frac{1}{2}, 0 )$ & $( \frac{1}{2}, \frac{1}{2}, 1 )$ \\
			$( \frac{5}{2}, \frac{1}{2}, \frac{1}{2} )$ & $( \frac{1}{2}, \frac{1}{2}, \frac{1}{2} )$ \\
			$( \frac{5}{2}, 1, \frac{1}{2} )$ & $( \frac{1}{2}, 0, \frac{1}{2} )$ \\
			$( \frac{5}{2}, \frac{3}{2}, \frac{1}{2} )$ & $( \frac{1}{2}, -\frac{1}{2}, \frac{1}{2} )$ \\\hline 
\end{tabular}
\caption{The possible magnetic 
fluxes in Type B within the range of 
$-2\leq m_{Q_2}^i \leq 2$, $i=1,2,3$.
Possible permutations among the three 2-tori are omitted.
Certain types of sign flipping are also omitted.}
\label{table:B}
\end{center}
\end{table}

\clearpage

Under the constrained magnetic fluxes in 
Tables~\ref{table:A} and \ref{table:B}, 
we further search for the realistic three generations 
of $u_R^c$, $d_R^c$ and $e_R^c$ satisfying 
the $U(1)_Y$ massless conditions~(\ref{eq:massless1}), 
(\ref{eq:massless2}) as well as the SUSY 
conditions~(\ref{eq:SUSYcond}).~\footnote{Here we 
do not constrain the number of charged-leptons, $L$, 
because some of them may be identified as higgsino fields.} 
As a result, within the range of 
$-10\leq m_{u_{R_2}^{c\,a}}^i\leq 10$, there are three 
choices for the $U(1)$ fluxes as follows,
\begin{equation}
\begin{array}{ll}
{\rm Case\,I} & 
m_4^i=m_5^i=m_6^i=-m_7^i=-m_8^i=-m_9^i, \\
& 
m_{10}^i=m_{11}^i=-m_{12}^i=-m_{13}^i, \\
&
(m_{u_{R_2}^{c\,a}}, m_{d_{R_3}^{c\,a}}, m_{e_{R}^{c\,a}}, 
m_{n_2^{a}}) =(1,0,0,1),\quad (a=4,5,6),\\
&
(m_{u_{R_2}^{c\,b}}, m_{d_{R_3}^{c\,b}}, m_{e_{R}^{c\,b}}, 
m_{n_2^{b}}) =(0,1,1,0),\quad (b=7,8,9),\\
&
(m_{u_{R_2}^{c\,d}}, m_{d_{R_3}^{c\,d}}, m_{e_{R}^{c\,d}}, 
m_{n_2^{d}}) =(0,0,0,0),\quad (d=10,11,12,13),
\end{array}
\label{eq:caseI}
\end{equation}
,
\begin{equation}
\begin{array}{ll}
{\rm Case\,II} & 
m_4^i=-m_5^i, \\
& 
m_6^i=m_7^i=m_8^i=m_9^i=-m_{10}^i=-m_{11}^i
=-m_{12}^i=-m_{13}^i, \\
&
(m_{u_{R_2}^{c\,4}}, m_{d_{R_3}^{c\,4}}, m_{e_{R}^{c\,4}}, 
m_{n_2^{4}}) =(3,0,0,3),\\
&
(m_{u_{R_2}^{c\,5}}, m_{d_{R_3}^{c\,5}}, m_{e_{R}^{c\,5}}, 
m_{n_2^{5}}) =(0,3,3,0),\\
&
(m_{u_{R_2}^{c\,a}}, m_{d_{R_3}^{c\,a}}, m_{e_{R}^{c\,a}}, 
m_{n_2^{a}}) =(0,0,0,0),\quad (a=6,7,8,9,10,11,12,13).
\end{array}
\label{eq:caseII}
\end{equation}
and
\begin{equation}
\begin{array}{ll}
{\rm Case\,III} & 
m_4^i=-m_5^i, \qquad m_6^i=-m_7^i,\\
& 
m_8^i=m_9^i=m_{10}^i=-m_{11}^i=-m_{12}^i=-m_{13}^i, \\
&
(m_{u_{R_2}^{c\,4}}, m_{d_{R_3}^{c\,4}}, m_{e_{R}^{c\,4}}, 
m_{n_2^{4}}) =(2,0,0,2),\\
&
(m_{u_{R_2}^{c\,5}}, m_{d_{R_3}^{c\,5}}, m_{e_{R}^{c\,5}}, 
m_{n_2^{5}}) =(0,2,2,0),\\
&
(m_{u_{R_2}^{c\,6}}, m_{d_{R_3}^{c\,6}}, m_{e_{R}^{c\,6}}, 
m_{n_2^{6}}) =(1,0,0,1),\\
&
(m_{u_{R_2}^{c\,7}}, m_{d_{R_3}^{c\,7}}, m_{e_{R}^{c\,7}}, 
m_{n_2^{7}}) =(0,1,1,0),\\
&
(m_{u_{R_2}^{c\,a}}, m_{d_{R_3}^{c\,a}}, m_{e_{R}^{c\,a}}, 
m_{n_2^{a}}) =(0,0,0,0),\quad (a=8,9,10,11,12,13).
\end{array}
\label{eq:caseIII}
\end{equation}
\begin{table}
\begin{center}
\begin{tabular}{|c|c|c|c|c|} \hline 
$(m_1^1,m_1^2,m_1^3)$ & $(m_2^1,m_2^2,m_2^3)$ 
& $(m_3^1,m_3^2,m_3^3)$ & $(m_4^1,m_4^2,m_4^3)$ 
& $(m_{10}^1,m_{10}^2,m_{10}^3)$  
\\ \hline
$(\frac{3}{2},0,1)$ & $(\frac{1}{2},1,0)$ 
& $(0,0,0)$ & $(\frac{1}{2},-2,1)$ 
& $(\frac{1}{2},1,0)$
\\ \hline
 \end{tabular}
 \caption{The typical values of $U(1)$ fluxes 
in the model of type A and ``Case I'' given by 
Eqs.~(\ref{eq:case}) and (\ref{eq:caseI}).}
\label{table:AI1}
\end{center}
\end{table} 
\begin{table}
\begin{center}
\begin{tabular}{|c|c|} \hline 
$(Q_1,Q_2,L_1,L_2,u_{R_1}^c,d_{R_1}^c,d_{R_2}^c,n_1)$ 
& ($2,1,2,1,0,0,0,0$)
\\ \hline
$(L_3^4,L_4^4,u_{R_2}^{c\,4},d_{R_3}^{c\,4},e_{R_1}^{c\,4},n_2^4)$ 
& ($0,8,1,0,0,1$)
\\ 
$(L_3^5,L_4^5,u_{R_2}^{c\,5},d_{R_3}^{c\,5},e_{R_1}^{c\,5},n_2^5)$ 
& ($0,8,1,0,0,1$)
\\
$(L_3^6,L_4^6,u_{R_2}^{c\,6},d_{R_3}^{c\,6},e_{R_1}^{c\,6},n_2^6)$ 
& ($0,8,1,0,0,1$)
\\ \hline
$(L_3^7,L_4^7,u_{R_2}^{c\,7},d_{R_3}^{c\,7},e_{R_1}^{c\,7},n_2^7)$ 
& ($-8,0,0,1,1,0$)
\\ 
$(L_3^8,L_4^8,u_{R_2}^{c\,8},d_{R_3}^{c\,8},e_{R_1}^{c\,8},n_2^8)$ 
& ($-8,0,0,1,1,0$)
\\
$(L_3^9,L_4^9,u_{R_2}^{c\,9},d_{R_3}^{c\,9},e_{R_1}^{c\,9},n_2^9)$ 
& ($-8,0,0,1,1,0$)
\\ \hline
$(L_3^{10},L_4^{10},u_{R_2}^{c\,10},d_{R_3}^{c\,10},e_{R_1}^{c\,10},n_2^{10})$ 
& ($-1,-2,0,0,0,0$)
\\ 
$(L_3^{11},L_4^{11},u_{R_2}^{c\,11},d_{R_3}^{c\,11},e_{R_1}^{c\,11},n_2^{11})$ 
& ($-1,-2,0,0,0,0$)
\\ 
$(L_3^{12},L_4^{12},u_{R_2}^{c\,12},d_{R_3}^{c\,12},e_{R_1}^{c\,12},n_2^{12})$ 
& ($2,1,0,0,0,0$)
\\ 
$(L_3^{13},L_4^{13},u_{R_2}^{c\,13},d_{R_3}^{c\,13},e_{R_1}^{c\,13},n_2^{13})$ 
& ($2,1,0,0,0,0$)
\\ \hline
 \end{tabular}
 \caption{The number of generations for the 
representations defined in the model of 
type A and ``Case I'' given by 
Eqs.~(\ref{eq:case}) and (\ref{eq:caseI}).}
\label{table:AI2}
\end{center}
\end{table} 
In the case of Type A, only ``Case I" is allowed as 
the realistic three-generation models. The 
typical $U(1)$ fluxes and the number of generations 
of matters are given by Tables~\ref{table:AI1} 
and \ref{table:AI2}. Under the $U(1)$ gauge symmetries, 
the following Yukawa couplings of quarks and leptons 
are allowed in terms of the renormalizable operators, 
\begin{equation} 
\begin{array}{llll}
 (Q_1, {\bar L}_3^4, u_{R_2}^{c\,4}), &  
(Q_2, {\bar L}_4^4, u_{R_2}^{c\,4}), & 
 (L_1, {\bar L}_3^4, n_2^4), &
 (L_2, {\bar L}_4^4, n_2^4), \\ 
  (Q_1, {\bar L}_3^5, u_{R_2}^{c\,5}), &  
(Q_2, {\bar L}_4^5, u_{R_2}^{c\,5}), & 
 (L_1, {\bar L}_3^5, n_2^5), &
 (L_2, {\bar L}_4^5, n_2^5), \\ 
  (Q_1, {\bar L}_3^6, u_{R_2}^{c\,6}), &  
(Q_2, {\bar L}_4^6, u_{R_2}^{c\,6}), & 
 (L_1, {\bar L}_3^6, n_2^6), &
 (L_2, {\bar L}_4^6, n_2^6), \\ 
(Q_1, L_4^7, d_{R_3}^{c\,7}), &  
(Q_2, L_3^7, d_{R_3}^{c\,7}), & 
(L_1, L_4^7, e_{R_1}^{c\,7}), &
(L_2, L_3^7, e_{R_1}^{c\,7}), \\ 
(Q_1, L_4^8, d_{R_3}^{c\,8}), &  
(Q_2, L_3^8, d_{R_3}^{c\,8}), & 
(L_1, L_4^8, e_{R_1}^{c\,8}), &
(L_2, L_3^8, e_{R_1}^{c\,8}), \\
(Q_1, L_4^9, d_{R_3}^{c\,9}), &  
(Q_2, L_3^9, d_{R_3}^{c\,9}), & 
(L_1, L_4^9, e_{R_1}^{c\,9}), &
(L_2, L_3^9, e_{R_1}^{c\,9}).   
\end{array}
\end{equation} 
These include useful Yukawa couplings to give all of the quarks and leptons masses when ${\bar L}_3^a, {\bar L}_4^a, L_3^b, L_4^b$ with $a=4,5,6$ and $b=7,8,9$ are identified as Higgs doublets and 
${\bar L}^a_{3,4}$ denote conjugate representations of 
$L^a_{3,4}$. 

Next, we consider the case of Type B. As the 
supersymmetric three-generation models, both 
``Case I" and ``Case II" are allowed and they are 
then categorized as the four types of models, 
\begin{equation}
\begin{array}{l}
{\rm BI}:\, {\rm ``Case I"~in~type~B},\\ 
{\rm BII}:\, {\rm ``Case II"~in~type~B~with}~m_{n_1}=0,\\ 
{\rm BIII}:\, {\rm ``Case II"~in~type~B~with}~m_{n_1}\neq 0,\\
{\rm BIV}:\, {\rm ``Case III"~in~type~B}. 
\label{eq:caseB}
\end{array}
\end{equation} 
For each model, the typical $U(1)$ fluxes and the 
number of generations of matters are summarized 
in Tables~\ref{table:BI1}, 
\ref{table:BI2}, \ref{table:BII1}, \ref{table:BII2}, 
\ref{table:BIII1} and \ref{table:BIII2}. 
In the type BI model summarized in 
Tables~\ref{table:BI1} and \ref{table:BI2}, 
the following Yukawa couplings of quarks and leptons 
are allowed in terms of the renormalizable operators, 
\begin{equation} 
\begin{array}{llllll}
 (Q_1, {\bar L}_3^4, u_{R_2}^{c\,4}), &  
(Q_1, {\bar L}_3^5, u_{R_2}^{c\,5}), & 
(Q_1, {\bar L}_3^6, u_{R_2}^{c\,6}), & 
 (L_1, {\bar L}_3^4, n_2^4), &
 (L_1, {\bar L}_3^5, n_2^5), &
 (L_1, {\bar L}_3^6, n_2^6), \\ 
(Q_1, L_4^7, d_{R_3}^{c\,7}), &  
(Q_1, L_4^8, d_{R_3}^{c\,8}), &
(Q_1, L_4^9, d_{R_3}^{c\,9}), & 
(L_1, L_4^7, e_{R_1}^{c\,7}), &
(L_1, L_4^8, e_{R_1}^{c\,8}), &
(L_1, L_4^9, e_{R_1}^{c\,9}).    
\end{array}
\end{equation} 
These also include useful Yukawa couplings when ${\bar L}_3^a, L_4^b$ 
with $a=4,5,6$ and $b=7,8,9$ are identified as Higgs doublets.
In both type BII and type BIII models summarized in 
Tables~\ref{table:BII1}, \ref{table:BII2}, \ref{table:BIII1} 
and \ref{table:BIII2}, 
the useful Yukawa couplings of quarks and leptons 
are allowed in terms of the renormalizable operators, 
\begin{equation} 
\begin{array}{llll}
 (Q_1, {\bar L}_3^4, u_{R_2}^{c\,4}), &  
(Q_1, L_4^5, d_{R_3}^{c\,5}), & 
 (L_1, {\bar L}_3^4, n_2^4), &
(L_1, L_4^5, e_{R_1}^{c\,5}), 
\end{array}
\end{equation} 
where ${\bar L}_3^4, L_4^5$ are identified 
as  Higgs doublets. 
Finally, in type BIV model summarized in Tables~\ref{table:BIV1} and \ref{table:BIV2}, 
the useful Yukawa couplings of quarks and leptons 
are allowed in terms of the renormalizable operators, 
\begin{equation} 
\begin{array}{llll}
 (Q_1, {\bar L}_3^4, u_{R_2}^{c\,4}), &  
(Q_1, {\bar L}_3^6, u_{R_2}^{c\,6}), & 
 (L_1, {\bar L}_3^4, n_2^4), &
 (L_1, {\bar L}_3^6, n_2^6), \\ 
(Q_1, L_4^5, d_{R_3}^{c\,5}), &  
(Q_1, L_4^7, d_{R_3}^{c\,7}), &
(L_1, L_4^5, e_{R_1}^{c\,5}), &
(L_1, L_4^, e_{R_1}^{c\,7}),   
\end{array}
\end{equation} 
where ${\bar L}_3^{4,6}, L_4^{5,7}$ are identified as Higgs doublets.

\begin{table}
\begin{center}
\begin{tabular}{|c|c|c|c|c|} \hline 
$(m_1^1,m_1^2,m_1^3)$ & $(m_2^1,m_2^2,m_2^3)$ 
& $(m_3^1,m_3^2,m_3^3)$ & $(m_4^1,m_4^2,m_4^3)$ 
& $(m_{10}^1,m_{10}^2,m_{10}^3)$  
\\ \hline
$(1,0,\frac{1}{2})$ & $(2,1,\frac{1}{2})$ 
& $(0,0,0)$ & $(-1,-2,\frac{1}{2})$ 
& $(0,1,-\frac{1}{2})$
\\ \hline
 \end{tabular}
 \caption{The typical values of $U(1)$ fluxes 
in the type BI model given by 
Eq.~(\ref{eq:caseB}).}
\label{table:BI1}
\end{center}
\end{table} 
\begin{table}
\begin{center}
\begin{tabular}{|c|c|} \hline 
$(Q_1,Q_2,L_1,L_2,u_{R_1}^c,d_{R_1}^c,d_{R_2}^c,n_1)$ 
& ($3,0,3,0,0,8,-8,0$)
\\ \hline
$(L_3^4,L_4^4,u_{R_2}^{c\,4},d_{R_3}^{c\,4},e_{R_1}^{c\,4},n_2^4)$ 
& ($0,0,1,0,0,1$)
\\ 
$(L_3^5,L_4^5,u_{R_2}^{c\,5},d_{R_3}^{c\,5},e_{R_1}^{c\,5},n_2^5)$ 
& ($0,0,1,0,0,1$)
\\
$(L_3^6,L_4^6,u_{R_2}^{c\,6},d_{R_3}^{c\,6},e_{R_1}^{c\,6},n_2^6)$ 
& ($0,0,1,0,0,1$)
\\ \hline
$(L_3^7,L_4^7,u_{R_2}^{c\,7},d_{R_3}^{c\,7},e_{R_1}^{c\,7},n_2^7)$ 
& ($0,0,0,1,1,0$)
\\ 
$(L_3^8,L_4^8,u_{R_2}^{c\,8},d_{R_3}^{c\,8},e_{R_1}^{c\,8},n_2^8)$ 
& ($0,0,0,1,1,0$)
\\
$(L_3^9,L_4^9,u_{R_2}^{c\,9},d_{R_3}^{c\,9},e_{R_1}^{c\,9},n_2^9)$ 
& ($0,0,0,1,1,0$)
\\ \hline
$(L_3^{10},L_4^{10},u_{R_2}^{c\,10},d_{R_3}^{c\,10},e_{R_1}^{c\,10},n_2^{10})$ 
& ($-1,0,0,0,0,0$)
\\ 
$(L_3^{11},L_4^{11},u_{R_2}^{c\,11},d_{R_3}^{c\,11},e_{R_1}^{c\,11},n_2^{11})$ 
& ($-1,0,0,0,0,0$)
\\ 
$(L_3^{12},L_4^{12},u_{R_2}^{c\,12},d_{R_3}^{c\,12},e_{R_1}^{c\,12},n_2^{12})$ 
& ($0,1,0,0,0,0$)
\\ 
$(L_3^{13},L_4^{13},u_{R_2}^{c\,13},d_{R_3}^{c\,13},e_{R_1}^{c\,13},n_2^{13})$ 
& ($0,1,0,0,0,0$)
\\ \hline
 \end{tabular}
 \caption{The number of generations for the 
representations in the type BI model 
given by Eq.~(\ref{eq:caseB}).}
\label{table:BI2}
\end{center}
\end{table} 
\begin{table}
\begin{center}
\begin{tabular}{|c|c|c|c|c|} \hline 
$(m_1^1,m_1^2,m_1^3)$ & $(m_2^1,m_2^2,m_2^3)$ 
& $(m_3^1,m_3^2,m_3^3)$ & $(m_4^1,m_4^2,m_4^3)$ 
& $(m_6^1,m_6^2,m_6^3)$  
\\ \hline
$(\frac{5}{2},0,\frac{1}{2})$ & $(\frac{1}{2},1,\frac{1}{2})$ 
& $(0,0,0)$ & $(-\frac{3}{2},2,\frac{1}{2})$ 
& $(\frac{9}{2},-1,\frac{1}{2})$
\\ \hline
 \end{tabular}
 \caption{The typical values of $U(1)$ fluxes 
in the type BII model given by Eq.~(\ref{eq:caseB}).}
\label{table:BII1}
\end{center}
\end{table} 
\begin{table}
\begin{center}
\begin{tabular}{|c|c|} \hline 
$(Q_1,Q_2,L_1,L_2,u_{R_1}^c,d_{R_1}^c,d_{R_2}^c,n_1)$ 
& ($3,0,3,0,0,2,-2,0$)
\\ \hline
$(L_3^4,L_4^4,u_{R_2}^{c\,4},d_{R_3}^{c\,4},e_{R_1}^{c\,4},n_2^4)$ 
& ($0,-2,3,0,0,3$)
\\ \hline
$(L_3^5,L_4^5,u_{R_2}^{c\,5},d_{R_3}^{c\,5},e_{R_1}^{c\,5},n_2^5)$ 
& ($2,0,0,3,3,0$)
\\ \hline
$(L_3^6,L_4^6,u_{R_2}^{c\,6},d_{R_3}^{c\,6},e_{R_1}^{c\,6},n_2^6)$ 
& ($0,7,0,0,0,0$)
\\ 
$(L_3^7,L_4^7,u_{R_2}^{c\,7},d_{R_3}^{c\,7},e_{R_1}^{c\,7},n_2^7)$ 
& ($0,7,0,0,0,0$)
\\ 
$(L_3^8,L_4^8,u_{R_2}^{c\,8},d_{R_3}^{c\,8},e_{R_1}^{c\,8},n_2^8)$ 
& ($0,7,0,0,0,0$)
\\
$(L_3^9,L_4^9,u_{R_2}^{c\,9},d_{R_3}^{c\,9},e_{R_1}^{c\,9},n_2^9)$ 
& ($0,7,0,0,0,0$)
\\ \hline
$(L_3^{10},L_4^{10},u_{R_2}^{c\,10},d_{R_3}^{c\,10},e_{R_1}^{c\,10},n_2^{10})$ 
& ($-7,0,0,0,0,0$)
\\ 
$(L_3^{11},L_4^{11},u_{R_2}^{c\,11},d_{R_3}^{c\,11},e_{R_1}^{c\,11},n_2^{11})$ 
& ($-7,0,0,0,0,0$)
\\ 
$(L_3^{12},L_4^{12},u_{R_2}^{c\,12},d_{R_3}^{c\,12},e_{R_1}^{c\,12},n_2^{12})$ 
& ($-7,0,0,0,0,0$)
\\ 
$(L_3^{13},L_4^{13},u_{R_2}^{c\,13},d_{R_3}^{c\,13},e_{R_1}^{c\,13},n_2^{13})$ 
& ($-7,0,0,0,0,0$)
\\ \hline
 \end{tabular}
 \caption{The number of generations for the 
representations in the type BII model 
given by Eq.~(\ref{eq:caseB}).}
\label{table:BII2}
\end{center}
\end{table} 
\begin{table}
\begin{center}
\begin{tabular}{|c|c|c|c|c|} \hline 
$(m_1^1,m_1^2,m_1^3)$ & $(m_2^1,m_2^2,m_2^3)$ 
& $(m_3^1,m_3^2,m_3^3)$ & $(m_4^1,m_4^2,m_4^3)$ 
& $(m_6^1,m_6^2,m_6^3)$  
\\ \hline
$(1,-\frac{1}{2},\frac{1}{2})$ & $(2,\frac{3}{2},\frac{1}{2})$ 
& $(0,0,0)$ & $(-1,-\frac{9}{2},\frac{1}{2})$ 
& $(2,\frac{13}{2},-\frac{1}{2})$
\\ \hline
 \end{tabular}
 \caption{The typical values of $U(1)$ fluxes 
in the type BIII model given by Eq.~(\ref{eq:caseB}).}
\label{table:BIII1}
\end{center}
\end{table} 
\begin{table}
\begin{center}
\begin{tabular}{|c|c|} \hline 
$(Q_1,Q_2,L_1,L_2,u_{R_1}^c,d_{R_1}^c,d_{R_2}^c,n_1)$ 
& ($3,0,3,0,0,12,-12,2$)
\\ \hline
$(L_3^4,L_4^4,u_{R_2}^{c\,4},d_{R_3}^{c\,4},e_{R_1}^{c\,4},n_2^4)$ 
& ($0,0,3,0,0,3$)
\\ \hline
$(L_3^5,L_4^5,u_{R_2}^{c\,5},d_{R_3}^{c\,5},e_{R_1}^{c\,5},n_2^5)$ 
& ($0,0,0,3,3,0$)
\\ \hline
$(L_3^6,L_4^6,u_{R_2}^{c\,6},d_{R_3}^{c\,6},e_{R_1}^{c\,6},n_2^6)$ 
& ($7,0,0,0,0,0$)
\\ 
\vdots & \vdots 
\\ 
$(L_3^9,L_4^9,u_{R_2}^{c\,9},d_{R_3}^{c\,9},e_{R_1}^{c\,9},n_2^9)$ 
& ($7,0,0,0,0,0$) 
\\ \hline
$(L_3^{10},L_4^{10},u_{R_2}^{c\,10},d_{R_3}^{c\,10},e_{R_1}^{c\,10},n_2^{10})$ 
& ($0,-7,0,0,0,0$)
\\ 
\vdots & \vdots 
\\ 
$(L_3^{13},L_4^{13},u_{R_2}^{c\,13},d_{R_3}^{c\,13},e_{R_1}^{c\,13},n_2^{13})$ 
& ($0,-7,0,0,0,0$)
\\ \hline
 \end{tabular}
 \caption{The number of generations for the 
representations in the type BIII model given by 
Eq.~(\ref{eq:caseB}).}
\label{table:BIII2}
\end{center}
\end{table} 
\clearpage
\begin{table}
\begin{center}
\begin{tabular}{|c|c|c|c|c|c|} \hline 
$(m_1^1,m_1^2,m_1^3)$ & $(m_2^1,m_2^2,m_2^3)$ 
& $(m_3^1,m_3^2,m_3^3)$ & $(m_4^1,m_4^2,m_4^3)$ 
& $(m_6^1,m_6^2,m_6^3)$  & $(m_8^1,m_8^2,m_8^3)$  
\\ \hline
$(\frac{5}{2},\frac{1}{2},\frac{1}{2})$ & $(\frac{1}{2},\frac{1}{2},\frac{1}{2})$ 
& $(0,0,0)$ & $(-\frac{5}{2},\frac{1}{2},\frac{1}{2})$ 
& $(-\frac{3}{2},\frac{1}{2},\frac{1}{2})$
& $(-\frac{1}{2},\frac{1}{2},\frac{1}{2})$
\\ \hline
 \end{tabular}
 \caption{The typical values of $U(1)$ fluxes 
in the type BIV model given by Eq.~(\ref{eq:caseB}).}
\label{table:BIV1}
\end{center}
\end{table} 
\begin{table}
\begin{center}
\begin{tabular}{|c|c|} \hline 
$(Q_1,Q_2,L_1,L_2,u_{R_1}^c,d_{R_1}^c,d_{R_2}^c,n_1)$ 
& ($3,0,3,0,0,1,-1,5$)
\\ \hline
$(L_3^4,L_4^4,u_{R_2}^{c\,4},d_{R_3}^{c\,4},e_{R_1}^{c\,4},n_2^4)$ 
& ($0,0,2,0,0,2$)
\\
$(L_3^5,L_4^5,u_{R_2}^{c\,5},d_{R_3}^{c\,5},e_{R_1}^{c\,5},n_2^5)$ 
& ($0,0,0,2,2,0$)
\\ \hline
$(L_3^6,L_4^6,u_{R_2}^{c\,6},d_{R_3}^{c\,6},e_{R_1}^{c\,6},n_2^6)$ 
& ($0,-1,1,0,0,1$)
\\ 
$(L_3^7,L_4^7,u_{R_2}^{c\,7},d_{R_3}^{c\,7},e_{R_1}^{c\,7},n_2^7)$ 
& ($0,-1,1,0,0,1$)
\\ \hline
$(L_3^8,L_4^8,u_{R_2}^{c\,8},d_{R_3}^{c\,8},e_{R_1}^{c\,8},n_2^8)$ 
& ($0,-2,0,0,0,0$) 
\\ 
$(L_3^9,L_4^9,u_{R_2}^{c\,9},d_{R_3}^{c\,9},e_{R_1}^{c\,9},n_2^9)$ 
& ($0,-2,0,0,0,0$) 
\\ 
$(L_3^{10},L_4^{10},u_{R_2}^{c\,10},d_{R_3}^{c\,10},e_{R_1}^{c\,10},n_2^{10})$ 
& ($0,-2,0,0,0,0$)
\\ \hline
$(L_3^{11},L_4^{11},u_{R_2}^{c\,11},d_{R_3}^{c\,11},e_{R_1}^{c\,11},n_2^{11})$ 
& ($2,0,0,0,0,0$)
\\ 
$(L_3^{12},L_4^{12},u_{R_2}^{c\,12},d_{R_3}^{c\,12},e_{R_1}^{c\,12},n_2^{12})$ 
& ($2,0,0,0,0,0$)
\\ 
$(L_3^{13},L_4^{13},u_{R_2}^{c\,13},d_{R_3}^{c\,13},e_{R_1}^{c\,13},n_2^{13})$ 
& ($2,0,0,0,0,0$)
\\ \hline
 \end{tabular}
 \caption{The number of generations for the 
representations in the type BIV model given by 
Eq.~(\ref{eq:caseB}).}
\label{table:BIV2}
\end{center}
\end{table} 

Note that in our models, 
the consistency conditions given by Eq.~(\ref{eq:tad2}) are not satisfied without introducing the heterotic five-branes. In this case, 
we have to take care of the Witten anomaly~\cite{Witten:1982fp,Witten:1985xe} 
on the heterotic five-branes with $Sp(2N)$ gauge groups 
which is the case that 
the number of heterotic five-branes is $N$. 
In order to avoid the Witten anomaly, the number of chiral fermions 
under the fundamental representations of $Sp(2N)$ are 
even~\cite{Witten:1982fp,Witten:1985xe}. 
These fundamental representations of $(32, 2N)$ 
under $SO(32)\otimes Sp(2N)$ can be read 
in the type I string with D$5$-and D$9$-brane system which 
is expected as the S-dual of the $SO(32)$ heterotic string. 
The generations of 
the chiral fermions included in $(12, 2N)$ under 
$SO(12) \otimes Sp(2N)$ and $(20, 2N)$ under 
$SO(20) \otimes Sp(2N)$ are determined by 
$\pm \prod_{i=1}^3m_a^{(i)}$ for $a=1,2,4,\cdots,13$, 
in the case $m_3^{(i)}=0$ with $i=1,2,3$. 
In our most supersymmetric models, the chiral 
fermions arise from $(32, 2N)$ under $SO(32)\otimes Sp(2N)$. Thus we require the non-trivial mechanism to obtain the even number of chiral fermions such as $U(1)$ fluxes on the heterotic 
five-branes in order to avoid the Witten anomaly. 

Finally we comment on the gauge enhancements induced by vanishing fluxes. 
In this paper, we focus on the case $m_3^i=0$, $i=1,2,3$ in the light of $U(1)_Y$ massless 
conditions given by Eqs.~(\ref{eq:massless1}) 
and (\ref{eq:massless2}). These vanishing 
fluxes cause the gauge enhancement, 
$SU(3)_C \times U(1)_3 \rightarrow SU(4)$. 
Moreover it requires the Wilson-lines into the 
internal component of $U(1)_3$ to break 
down $SU(4)$ into $SU(3)$. 
Our models have other  gauge enhancements. 
The realistic three-generation 
models are summarized in three cases, ``Case I", ``Case II" and ``Case III" in Eqs.~(\ref{eq:caseI}),~(\ref{eq:caseII}) and~(\ref{eq:caseIII}), 
respectively. 
In both cases, most magnetic fluxes are related 
to each other due to the $U(1)_Y$ massless 
conditions given by Eqs.~(\ref{eq:massless1}) 
and (\ref{eq:massless2}). 
For example, the invariant simple roots under 
the existences of fluxes read
\begin{equation}
\begin{array}{ll}
{\rm Case\,I} & 
\alpha_1= (0,0,0,0,0,0;1,-1,0,\cdots, 0),\\
&\alpha_2= (0,0,0,0,0,0;0,1,-1,0,\cdots, 0),\\
&\alpha_3= (0,0,0,0,0,0;0,0,0,1,-1,0,\cdots, 0),\\
&\alpha_4= (0,0,0,0,0,0;0,0,0,0,1,-1,0,\cdots, 0),\\
&\alpha_5= (0,0,0,0,0,0;0,0,1,0,0,1,0,\cdots, 0),\\
\\
{\rm Case\,II} & 
\alpha_1= (0,0,0,0,0,0;1,1,0,\cdots, 0),
\\
{\rm Case\,III} & 
\alpha_1= (0,0,0,0,0,0;1,1,0,\cdots, 0),\\
&\alpha_2= (0,0,0,0,0,0;0,0,1,1,0,\cdots, 0),
\end{array}
\end{equation}
which implies the $SU(6)$, $SU(2)$ and $SU(2)\times SU(2)$ gauge symmetries, respectively. 
All of them include $SU(2)_R$. 
Furthermore, $SU(3)$ of $SU(6)$ is a flavor symmetry 
of right-handed matter fields, 
and the three  right-handed matter  generations is a triplet under 
$SU(3)$ flavor symmetry, while the left-handed matter fields are singlets.
We introduce Wilson lines to break theses symmetries.

\subsection{Three-generation models with K-theory constraints}
\label{subsec:threeK}
So far, we have not considered the so-called K-theory constraints which 
are formulated in the S-dual to the $SO(32)$ heterotic string theory, i.e., Type I 
string theory. 
In the $SO(32)$ heterotic string theory, the total number of magnetic fluxes is further constrained as 
\begin{align}
\sum_{a=1}^{13}m_a^i=0\,\,\,({\rm mod}\,2),
\label{eq:Kth}
\end{align}
for $i=1,2,3$, as stated in Ref.~\cite{Blumenhagen:2005ga}. 
Such a condition allows for the well-defined spinor representation of the gauge bundle, otherwise its 
wavefunction is not single-valued.

When we assume that the $SO(32)$ heterotic string theory on our gauge background is described as its S-dual theory, i.e., Type I string theory, 
the above condition~(\ref{eq:Kth}) may correspond to the K-theory constraints~\cite{Witten:1998cd} which cannot be classified in terms of a homology. 
These constraints can be understood by introducing all the possible probe D-branes~\cite{Uranga:2000xp}, and then they show the existence of several 
stable non-BPS branes with the discrete K-theory charge, i.e., $Z_2$-charge. 
In the case of $N$ stacks of heterotic five-brane with $Sp(2N)$ gauge group, 
they require the condition~(\ref{eq:Kth}) in order to avoid the Witten anomaly~\cite{Witten:1982fp,Witten:1985xe}.\footnote{In the heterotic string, the K-theory may be understood in terms of closed string 
tachyon~\cite{Garcia-Etxebarria:2014txa} based on supercritical string~\cite{Hellerman:2004zm}.} 
Furthermore, in type I string, the fractional fluxes are allowed due to multiple wrapping numbers of D-branes. 
Although such a degree of freedom is expected to appear in the heterotic string side, 
we do not consider these possibilities, which we leave for future works. 
Since all the models discussed in Sec.~\ref{subsec:three} do not satisfy the K-theory condition, 
in this section, we further search for the possibilities of three-generation models under these assumptions. 

First of all, in the light of $U(1)_Y$ massless condition, we impose the constraints for $U(1)$ fluxes as, 
\begin{align}
m_3^i=0,\,\,\,\,m_{a+3}^i=-m_{a+8}^i\,\,(a=1,2,3,4,5),
\label{eq:U(1)YK}
\end{align} 
with $i=1,2,3$, which simplify the K-theory condition as 
\begin{align}
\sum_{a=1}^{2}m_a^i=0\,\,\,({\rm mod}\,2).
\label{eq:Kth2}
\end{align}
From the fact that all the possible candidates for left-handed quarks $Q$ and charged leptons $L$ are involved in the adjoint representation of $SO(12)$, 
three generations of $Q$ and $L$ have to be realized from such a representation.  
Then, their fluxes are constrained as 
\begin{align}
&{\rm Type\,A}:\,(m_{Q_1},m_{Q_2})=(2,1),\quad
{\rm Type\,A'}:\,(m_{Q_1},m_{Q_2})=(1,2),\nonumber\\
&{\rm Type\,B}:\,(m_{Q_1},m_{Q_2})=(3,0),\quad 
{\rm Type\,B'}:\,(m_{Q_1},m_{Q_2})=(0,3),
\label{eq:caseK}
\end{align}
where $m_{Q_{1,2}}^{i}$, and hereafter we focus on the case that the right-handed quarks $d_R^c$ are generated from the vector representation of $SO(12)$, for simplicity. 
In such cases, we find that only Type ${\rm B'}$  in Eq.~(\ref{eq:caseK}) satisfies the K-theory condition~(\ref{eq:Kth2}) and the SUSY condition~(\ref{eq:SUSYcond}) 
yielding three generations of $Q$ and $L$. The possible $U(1)_{1,2}$ fluxes are summarized in Tab.~\ref{table:B'}. 

\begin{table}[htb]
\begin{center}
\begin{tabular}{|c|c|} \hline
 $ \left( m^1_1, m^2_1,m^3_1 \right) $ & $ \left( m^1_2,m^2_2,m^3_2 \right) $ \\ \hline \hline
($ -\frac{5}{2} $, $ -\frac{3}{2} $, $ -\frac{1}{2} $) & ($ \frac{1}{2} $, $ -\frac{1}{2} $, $ \frac{1}{2} $) \\
($ -\frac{5}{2} $, $ -\frac{1}{2} $, $ -\frac{1}{2} $) & ($ \frac{1}{2} $, $ \frac{1}{2} $, $ \frac{1}{2} $) \\
($ -\frac{5}{2} $, $ -\frac{1}{2} $, $ \frac{1}{2} $) & ($ \frac{1}{2} $, $ \frac{1}{2} $, $ \frac{3}{2} $) \\
($ -\frac{3}{2} $, $ -\frac{3}{2} $, $ -\frac{3}{2} $) & ($ \frac{3}{2} $, $ -\frac{1}{2} $, $ -\frac{1}{2} $) \\
($ -\frac{3}{2} $, $ -\frac{3}{2} $, $ -\frac{1}{2} $) & ($ \frac{3}{2} $, $ -\frac{1}{2} $, $ \frac{1}{2} $) \\
($ -\frac{3}{2} $, $ -\frac{3}{2} $, $ \frac{1}{2} $) & ($ \frac{3}{2} $, $ -\frac{1}{2} $, $ \frac{3}{2} $) \\
($ -\frac{3}{2} $, $ -\frac{1}{2} $, $ -\frac{1}{2} $) & ($ \frac{3}{2} $, $ \frac{1}{2} $, $ \frac{1}{2} $) \\
($ -\frac{3}{2} $, $ -\frac{1}{2} $, $ \frac{1}{2} $) & ($ \frac{3}{2} $, $ \frac{1}{2} $, $ \frac{3}{2} $) \\
($ -\frac{3}{2} $, $ \frac{1}{2} $, $ \frac{1}{2} $) & ($ \frac{3}{2} $, $ \frac{3}{2} $, $ \frac{3}{2} $) \\
($ -\frac{1}{2} $, $ -\frac{3}{2} $, $ -\frac{1}{2} $) & ($ \frac{5}{2} $, $ -\frac{1}{2} $, $ \frac{1}{2} $) \\
($ -\frac{1}{2} $, $ -\frac{1}{2} $, $ -\frac{1}{2} $) & ($ \frac{5}{2} $, $ \frac{1}{2} $, $ \frac{1}{2} $) \\
($ -\frac{1}{2} $, $ -\frac{1}{2} $, $ \frac{1}{2} $) & ($ \frac{5}{2} $, $ \frac{1}{2} $, $ \frac{3}{2} $) 
\\ \hline 
\end{tabular}
\caption{The possible magnetic 
fluxes in ${\rm Type\,B'}$ within the range of 
$-2\leq m_{Q_{1}}^i \leq 2$ for $m_{Q_{2}}^i=0$ and $-2\leq m_{Q_{2}}^i \leq 2$ for $m_{Q_{1}}^i=0$, where $i=1,2,3$.}
\label{table:B'}
\end{center}
\end{table}

Next, we consider the remaining matter contents in the standard model, that is, $u_R^c$, $d_R^c$ and $e_R^c$. 
Among the constrained magnetic fluxes listed in Table~\ref{table:B'}, we further search for those yield three generations of $u_R^c$, $d_R^c$ and $e_R^c$, satisfying the $U(1)_Y$ 
massless condition~(\ref{eq:U(1)YK}) as well as the SUSY condition~(\ref{eq:SUSYcond}). Note that the K-theory condition is already satisfied under the constraints~(\ref{eq:U(1)YK}) and (\ref{eq:Kth2}). 
As a result, within the range of $-5\leq m_{u_{R_2}^{c\,a}}^i \leq 5$, there are three allowed choices for the $U(1)$ fluxes as follows,
\begin{equation}
\begin{array}{ll}
{\rm  ``Case\,I'"} & 
(m_{u_{R_2}^{c\,4}}, m_{u_{R_2}^{c\,5}},..., m_{u_{R_2}^{c\,13}}) =(1,1,1,0,0,0,0,0,0,0),\\
&
(m_{d_{R_2}^{c\,4}}, m_{d_{R_2}^{c\,5}},..., m_{d_{R_2}^{c\,13}}) =(0,0,0,0,0,1,1,1,0,0),
\end{array}
\label{eq:caseI'}
\end{equation}
\begin{equation}
\begin{array}{ll}
{\rm ``Case\,II'"} &  
(m_{u_{R_2}^{c\,4}}, m_{u_{R_2}^{c\,5}},..., m_{u_{R_2}^{c\,13}}) =(3,0,0,0,0,0,0,0,0,0),\\
&
(m_{d_{R_2}^{c\,4}}, m_{d_{R_2}^{c\,5}},..., m_{d_{R_2}^{c\,13}}) =(0,0,0,0,0,3,0,0,0,0),
\end{array}
\label{eq:caseII'}
\end{equation}
and 
\begin{equation}
\begin{array}{ll}
{\rm ``Case\,III'"} & 
(m_{u_{R_2}^{c\,4}}, m_{u_{R_2}^{c\,5}},..., m_{u_{R_2}^{c\,13}}) =(2,1,0,0,0,0,0,0,0,0),\\
&
(m_{d_{R_2}^{c\,4}}, m_{d_{R_2}^{c\,5}},..., m_{d_{R_2}^{c\,13}}) =(0,0,0,0,0,2,1,0,0,0).
\end{array}
\label{eq:caseIII'}
\end{equation}
For each model, the typical $U(1)$ fluxes and the number of generations of matters are summarized in Tables~\ref{table:I'1}, \ref{table:I'2}, \ref{table:II'1}, \ref{table:II'2}, \ref{table:III'1} and \ref{table:III'2}. 
In the ${\rm ``Case\,I'"}$ summarized in Tables~\ref{table:I'1} and \ref{table:I'2}, 
non-vanishing Yukawa coupling terms involving the following combinations of quarks and leptons 
are allowed as the renormalizable operators, 
\begin{equation} 
\begin{array}{llllll}
 (Q_2, {\bar L}_4^4, u_{R_2}^{c\,4}), &  
(Q_2, {\bar L}_4^5, u_{R_2}^{c\,5}), & 
(Q_2, {\bar L}_4^6, u_{R_2}^{c\,6}), & 
 (L_2, {\bar L}_4^4, n_2^4), &
 (L_2, {\bar L}_4^5, n_2^5), &
 (L_2, {\bar L}_4^6, n_2^6), \\ 
(Q_2, L_3^9, d_{R_3}^{c\,9}), &  
(Q_2, L_3^{10}, d_{R_3}^{c\,10}), &
(Q_2, L_3^{11}, d_{R_3}^{c\,11}), & 
(L_2, L_3^9, e_{R_1}^{c\,9}), &
(L_2, L_3^{10}, e_{R_1}^{c\,10}), &
(L_2, L_3^{11}, e_{R_1}^{c\,11}).    
\end{array}
\end{equation} 
These include useful Yukawa couplings to give masses of all the quarks and leptons 
when ${\bar L}_4^a, L_3^b$ with $a=4,5,6$ and $b=9,10,11$ are identified as Higgs doublets. 

As for the ${\rm ``Case\,II'"}$ summarized in Tables~\ref{table:II'1} and \ref{table:II'2}, 
the following combinations of quarks and leptons have renormalizable Yukawa coupling, 
\begin{equation} 
\begin{array}{llll}
 (Q_2, {\bar L}_4^4, u_{R_2}^{c\,4}), &  
(Q_2, L_3^9, d_{R_3}^{c\,9}), & 
(L_2, {\bar L}_4^4, n_2^4), &
(L_2, L_3^9, e_{R_1}^{c\,9}), 
\end{array}
\end{equation} 
where ${\bar L}_4^4, L_3^9$ are identified as  Higgs doublets in order to be phenomenologically viable. 

Next, in the ${\rm ``Case\,III'"}$ summarized in Tables~\ref{table:III'1} and \ref{table:III'2}, 
the renormalizable Yukawa couplings are allowed for the following combinations of quarks and leptons, 
\begin{equation} 
\begin{array}{llll}
 (Q_2, {\bar L}_4^4, u_{R_2}^{c\,4}), &  
(Q_2, {\bar L}_4^5, u_{R_2}^{c\,5}), & 
(L_2, {\bar L}_4^4, n_2^4), &
(L_2, {\bar L}_4^5, n_{2}^{5}), \\
(Q_2, L_3^9, d_{R_3}^{c\,9}), &  
(Q_2, L_3^{10}, d_{R_3}^{c\,10}), &  
(L_2, L_3^9, e_{R_1}^{c\,9}), &
(L_2, L_3^{10}, e_{R_1}^{c\,10}), 
\end{array}
\end{equation} 
where ${\bar L}_4^{4,5}, L_3^{9,10}$ can be identified as Higgs doublets. 
Note that in the same way as in Sec.~\ref{subsec:three}, the consistency conditions given by Eq.~(\ref{eq:tad2}) are not satisfied without introducing the heterotic five-branes, in the supersymmetric case. 

Finally we comment on the gauge enhancements induced by vanishing fluxes. 
As discussed in Sec.~\ref{subsec:three}, vanishing $U(1)_3$ fluxes require the existence of 
Wilson-lines for the internal component of $U(1)_3$ to break $SU(4)$ down to $SU(3)$. 
There are other gauge enhancements in three realistic models, ${\rm ``Case\,I'"}$, ${\rm ``Case\,II'"}$ and ${\rm ``Case\,III'"}$, where 
most magnetic fluxes are related to each other due to the $U(1)_Y$ massless conditions~(\ref{eq:massless1}), (\ref{eq:massless2}) and the K-theory condition~(\ref{eq:Kth}).  
For example, there are invariant simple roots under the existences of fluxes such as $SU(6)$, $SU(2)$ and $SU(2)\times SU(2)$ gauge symmetries for the ${\rm ``Case\,I'"}$, 
${\rm ``Case\,II'"}$ and ${\rm ``Case\,III'"}$, respectively. 
We introduce Wilson-lines to break these gauge symmetries.

\begin{table}
\begin{center}
\begin{tabular}{|c|c|c|c|c|} \hline 
$(m_1^1,m_1^2,m_1^3)$ & $(m_2^1,m_2^2,m_2^3)$ 
& $(m_3^1,m_3^2,m_3^3)$ & $(m_4^1,m_4^2,m_4^3)$ 
& $(m_7^1,m_7^2,m_7^3)$  
\\ \hline
$(-\frac{3}{2},-\frac{1}{2},-\frac{1}{2})$ & $(\frac{3}{2},\frac{1}{2},\frac{1}{2})$ 
& $(0,0,0)$ & $(-\frac{1}{2},\frac{1}{2},-\frac{3}{2})$ 
& $(\frac{1}{2},\frac{1}{2},-\frac{1}{2})$
\\ \hline
 \end{tabular}
 \caption{The typical values of $U(1)$ fluxes 
in the ${\rm ``Case\,I'"}$ given by Eq.~(\ref{eq:caseI'}). The other $U(1)$ fluxes are constrained to be $m_4^i=m_5^i=m_6^i=-m_9^i=-m_{10}^i=-m_{11}^i$ and $m_{7}^i=m_{8}^i=-m_{12}^i=-m_{13}^i$ with $i=1,2,3$.}
\label{table:I'1}
\end{center}
\end{table} 
\begin{table}
\begin{center}
\begin{tabular}{|c|c|} \hline 
$(Q_1,Q_2,L_1,L_2,u_{R_1}^c,d_{R_1}^c,d_{R_2}^c,n_1)$ 
& ($0,3,0,3,0,3,-3,3$)
\\ \hline
$(L_3^4,L_4^4,u_{R_2}^{c\,4},d_{R_3}^{c\,4},e_{R_1}^{c\,4},n_2^4)$ 
& ($1,0,1,0,0,1$)
\\ 
$(L_3^5,L_4^5,u_{R_2}^{c\,5},d_{R_3}^{c\,5},e_{R_1}^{c\,5},n_2^5)$ 
& ($1,0,1,0,0,1$)
\\
$(L_3^6,L_4^6,u_{R_2}^{c\,6},d_{R_3}^{c\,6},e_{R_1}^{c\,6},n_2^6)$ 
& ($1,0,1,0,0,1$)
\\ \hline
$(L_3^7,L_4^7,u_{R_2}^{c\,7},d_{R_3}^{c\,7},e_{R_1}^{c\,7},n_2^7)$ 
& ($0,0,0,0,0,0$)
\\ 
$(L_3^8,L_4^8,u_{R_2}^{c\,8},d_{R_3}^{c\,8},e_{R_1}^{c\,8},n_2^8)$ 
& ($0,0,0,0,0,0$)
\\ \hline
$(L_3^9,L_4^9,u_{R_2}^{c\,9},d_{R_3}^{c\,9},e_{R_1}^{c\,9},n_2^9)$ 
& ($0,-1,0,1,1,0$)
\\ 
$(L_3^{10},L_4^{10},u_{R_2}^{c\,10},d_{R_3}^{c\,10},e_{R_1}^{c\,10},n_2^{10})$ 
& ($0,-1,0,1,1,0$)
\\ 
$(L_3^{11},L_4^{11},u_{R_2}^{c\,11},d_{R_3}^{c\,11},e_{R_1}^{c\,11},n_2^{11})$ 
& ($0,-1,0,1,1,0$)
\\ \hline
$(L_3^{12},L_4^{12},u_{R_2}^{c\,12},d_{R_3}^{c\,12},e_{R_1}^{c\,12},n_2^{12})$ 
& ($0,0,0,0,0,0$)
\\ 
$(L_3^{13},L_4^{13},u_{R_2}^{c\,13},d_{R_3}^{c\,13},e_{R_1}^{c\,13},n_2^{13})$ 
& ($0,0,0,0,0,0$)
\\ \hline
 \end{tabular}
 \caption{The number of generations for the 
representations in the ${\rm ``Case\,I'"}$ given by Eq.~(\ref{eq:caseI'}).}
\label{table:I'2}
\end{center}
\end{table} 
\begin{table}
\begin{center}
\begin{tabular}{|c|c|c|c|c|} \hline 
$(m_1^1,m_1^2,m_1^3)$ & $(m_2^1,m_2^2,m_2^3)$ 
& $(m_3^1,m_3^2,m_3^3)$ & $(m_4^1,m_4^2,m_4^3)$ 
& $(m_5^1,m_5^2,m_5^3)$  
\\ \hline
$(-\frac{1}{2},-\frac{1}{2},-\frac{1}{2})$ & $(\frac{5}{2},\frac{1}{2},\frac{1}{2})$ 
& $(0,0,0)$ & $(-\frac{11}{2},\frac{1}{2},\frac{1}{2})$ 
& $(\frac{5}{2},-\frac{1}{2},-\frac{1}{2})$
\\ \hline
 \end{tabular}
 \caption{The typical values of $U(1)$ fluxes in the ${\rm ``Case\,II'"}$ given by Eq.~(\ref{eq:caseII'}). 
The other $U(1)$ fluxes are constrained to be $m_4^i=-m_9^i$ and $m_5^i=m_6^i=m_7^i=m_8^i=-m_{10}^i=-m_{11}^i=-m_{12}^i=-m_{13}^i$ with $i=1,2,3$.}
\label{table:II'1}
\end{center}
\end{table} 
\begin{table}
\begin{center}
\begin{tabular}{|c|c|} \hline 
$(Q_1,Q_2,L_1,L_2,u_{R_1}^c,d_{R_1}^c,d_{R_2}^c,n_1)$ 
& ($0,3,0,3,0,5,-5,1$)
\\ \hline
$(L_3^4,L_4^4,u_{R_2}^{c\,4},d_{R_3}^{c\,4},e_{R_1}^{c\,4},n_2^4)$ 
& ($5,0,3,0,0,3$)
\\ \hline
$(L_3^5,L_4^5,u_{R_2}^{c\,5},d_{R_3}^{c\,5},e_{R_1}^{c\,5},n_2^5)$ 
& ($0,-2,0,0,0,0$)
\\
$(L_3^6,L_4^6,u_{R_2}^{c\,6},d_{R_3}^{c\,6},e_{R_1}^{c\,6},n_2^6)$ 
& ($0,-2,0,0,0,0$)
\\ 
$(L_3^7,L_4^7,u_{R_2}^{c\,7},d_{R_3}^{c\,7},e_{R_1}^{c\,7},n_2^7)$ 
& ($0,-2,0,0,0,0$)
\\ 
$(L_3^8,L_4^8,u_{R_2}^{c\,8},d_{R_3}^{c\,8},e_{R_1}^{c\,8},n_2^8)$ 
& ($0,-2,0,0,0,0$)
\\ \hline
$(L_3^9,L_4^9,u_{R_2}^{c\,9},d_{R_3}^{c\,9},e_{R_1}^{c\,9},n_2^9)$ 
& ($0,-5,0,3,3,0$)
\\ \hline
$(L_3^{10},L_4^{10},u_{R_2}^{c\,10},d_{R_3}^{c\,10},e_{R_1}^{c\,10},n_2^{10})$ 
& ($2,0,0,0,0,0$)
\\ 
$(L_3^{11},L_4^{11},u_{R_2}^{c\,11},d_{R_3}^{c\,11},e_{R_1}^{c\,11},n_2^{11})$ 
& ($2,0,0,0,0,0$)
\\ 
$(L_3^{12},L_4^{12},u_{R_2}^{c\,12},d_{R_3}^{c\,12},e_{R_1}^{c\,12},n_2^{12})$ 
& ($2,0,0,0,0,0$)
\\ 
$(L_3^{13},L_4^{13},u_{R_2}^{c\,13},d_{R_3}^{c\,13},e_{R_1}^{c\,13},n_2^{13})$ 
& ($2,0,0,0,0,0$)
\\ \hline
 \end{tabular}
 \caption{The number of generations for the 
representations in the ${\rm ``Case\,II'"}$ given by Eq.~(\ref{eq:caseII'}).}
\label{table:II'2}
\end{center}
\end{table} 
\begin{table}
\begin{center}
\begin{tabular}{|c|c|c|c|c|c|} \hline 
$(m_1^1,m_1^2,m_1^3)$ & $(m_2^1,m_2^2,m_2^3)$ 
& $(m_3^1,m_3^2,m_3^3)$ & $(m_4^1,m_4^2,m_4^3)$ 
& $(m_5^1,m_5^2,m_5^3)$  & $(m_6^1,m_6^2,m_6^3)$  
\\ \hline
$(-\frac{3}{2},-\frac{1}{2},-\frac{1}{2})$ & $(\frac{3}{2},\frac{1}{2},\frac{1}{2})$ 
& $(0,0,0)$ & $(-\frac{7}{2},\frac{1}{2},\frac{1}{2})$ 
& $(-\frac{5}{2},\frac{1}{2},\frac{1}{2})$ & $(\frac{3}{2},-\frac{1}{2},-\frac{1}{2})$
\\ \hline
 \end{tabular}
 \caption{The typical values of $U(1)$ fluxes in the ${\rm ``Case\,III'"}$ given by Eq.~(\ref{eq:caseIII'}). The other $U(1)$ fluxes are constrained to be $m_6^i=m_7^i=m_8^i=-m_{11}^i=-m_{12}^i=-m_{13}^i$ with $i=1,2,3$.}
\label{table:III'1}
\end{center}
\end{table} 
\begin{table}
\begin{center}
\begin{tabular}{|c|c|} \hline 
$(Q_1,Q_2,L_1,L_2,u_{R_1}^c,d_{R_1}^c,d_{R_2}^c,n_1)$ 
& ($0,3,0,3,0,3,-3,3$)
\\ \hline
$(L_3^4,L_4^4,u_{R_2}^{c\,4},d_{R_3}^{c\,4},e_{R_1}^{c\,4},n_2^4)$ 
& ($2,0,2,0,0,2$)
\\ 
$(L_3^5,L_4^5,u_{R_2}^{c\,5},d_{R_3}^{c\,5},e_{R_1}^{c\,5},n_2^5)$ 
& ($1,0,1,0,0,1$)
\\ \hline
$(L_3^6,L_4^6,u_{R_2}^{c\,6},d_{R_3}^{c\,6},e_{R_1}^{c\,6},n_2^6)$ 
& ($0,0,0,0,0,0$)
\\ 
$(L_3^7,L_4^7,u_{R_2}^{c\,7},d_{R_3}^{c\,7},e_{R_1}^{c\,7},n_2^7)$ 
& ($0,0,0,0,0,0$)
\\ 
$(L_3^8,L_4^8,u_{R_2}^{c\,8},d_{R_3}^{c\,8},e_{R_1}^{c\,8},n_2^8)$ 
& ($0,0,0,0,0,0$)
\\ \hline
$(L_3^9,L_4^9,u_{R_2}^{c\,9},d_{R_3}^{c\,9},e_{R_1}^{c\,9},n_2^9)$ 
& ($0,-2,0,2,2,0$)
\\ 
$(L_3^{10},L_4^{10},u_{R_2}^{c\,10},d_{R_3}^{c\,10},e_{R_1}^{c\,10},n_2^{10})$ 
& ($0,-1,0,1,1,0$)
\\ \hline
$(L_3^{11},L_4^{11},u_{R_2}^{c\,11},d_{R_3}^{c\,11},e_{R_1}^{c\,11},n_2^{11})$ 
& ($0,0,0,0,0,0$)
\\ 
$(L_3^{12},L_4^{12},u_{R_2}^{c\,12},d_{R_3}^{c\,12},e_{R_1}^{c\,12},n_2^{12})$ 
& ($0,0,0,0,0,0$)
\\ 
$(L_3^{13},L_4^{13},u_{R_2}^{c\,13},d_{R_3}^{c\,13},e_{R_1}^{c\,13},n_2^{13})$ 
& ($0,0,0,0,0,0$)
\\ \hline
 \end{tabular}
 \caption{The number of generations for the 
representations in the ${\rm ``Case\,III'"}$ given by Eq.~(\ref{eq:caseIII'}).}
\label{table:III'2}
\end{center}
\end{table} 

\clearpage
\section{Conclusion}
\label{sec:con}
In this paper, we have derived the realistic 
standard model gauge groups from the 
framework of $SO(32)$ heterotic string 
theory on three factorizable 2-tori with magnetic fluxes. 
Introducing magnetic fluxes as well as Wilson lines 
into Cartan directions of $SO(32)$ 
break $SO(32)$ to $SU(3) \times SU(2) \times U(1)_Y$ 
and extra symmetries.
These $U(1)$ fluxes also lead to chiral 
fermions in the four dimensions if and only if 
the fluxes insert into all the three 2-tori. 
At the same time, the 
generations of chiral matters are determined 
by the numbers of fluxes.
We have derived three chiral generations of 
quarks and leptons.
Our models also include Higgs fields, 
which have Yukawa couplings to quarks and leptons 
at tree level.

Possible configurations of magnetic fluxes 
are severely constrained 
by the massless condition of $U(1)_Y$ hypercharge 
gauge boson and the consistency condition of heterotic string theory. 
It is remarkable that in general, 
the ten-dimensional Green-Schwarz term 
induces the Stueckelberg couplings to 
multiple $U(1)$ gauge bosons which might lead 
to the mass term of $U(1)_Y$ hypercharge gauge 
boson. In this respect, the numbers of fluxes 
have been constrained by the massless condition of $U(1)_Y$ gauge boson. 
Since the torus is flat, our models requires the existence of 
heterotic five-branes in order to satisfy 
the consistency conditions without introducing 
the extra Stueckelberg couplings to $U(1)$ gauge boson, in contrast to the $E_8\times E_8$ 
heterotic string theory. 
At that time, the Witten anomaly cancellation constrains the 
number of $U(1)$ fluxes due to the nature of symplectic groups 
on the heterotic five-branes. 
In fact, the chiral fermions under the fundamental representation 
of symplectic gauge groups do not arise in the parts of our 
models, whereas the other parts of our models requires the 
non-trivial mechanisms such as $U(1)$ fluxes on heterotic 
five-branes to cause even number of these chiral 
fermions to avoid the Witten anomaly. 
We listed  supersymmetric three-generation standard models 
with massless $U(1)_Y$ gauge bosons  
and desirable Yukawa couplings of quarks, leptons and Higgs. 
The detailed phenomenological analysis of our models such as mass matrices would be studied in a separate work and the detail of 
this paper is applicable in the framework of type I string. 

The unbroken gauge sector in our models has ${\cal N}=4$ supersymmetry, that is, 
three adjoint scalar fields and four types of gaugino fields. 
However, the existence of (anti-)heterotic five-branes 
would lead to the breaking of (all) partial breaking of 
supersymmetry in our model. 
Orbifolding would be useful to reduce ${\cal N}=4$ 
supersymmetry to ${\cal N}=1$. 
Zero-mode wavefuctions have been also studied on
orbifolds with magnetic fluxes \cite{Abe:2008fi,Abe:2013bca}.
Such extensions would be also interesting.

\subsection*{Acknowledgement}
H.~O. would like to thank T.~Higaki for useful discussions and 
comments. 
H.~A. was supported in
part by the Grant-in-Aid for Scientific Research No. 25800158 from the
Ministry of Education,
Culture, Sports, Science and Technology (MEXT) in Japan. T.~K. was
supported in part by
the Grant-in-Aid for Scientific Research No.~25400252 and No.~26247042 from the MEXT in
Japan.
H.~O. was supported in part by a Grant-in-Aid for JSPS Fellows 
No. 26-7296.

\appendix

\section{Normalization of the $SO(32)$ gauge group}
\label{app:normalization}
In this appendix, we show the normalization of 
Abelian gauge groups embedded in $SO(32)$ 
gauge group. (For more details, see 
Refs.~\cite{Polchinsky,Font:1989aj,Blumenhagen}.) 
First, we comment on the normalization about the non-Abelian gauge groups 
in $SO(32)$. 
The sum of each Coxeter labels associated with the 
simple roots of $SO(32)$ are called as the Coxeter 
number $h(g)$ which is related to the 
quadratic Casimir via the following relation,
\begin{align}
\sum_{c,d}f^{acd}f^{bcd}= h(g)\psi^2 \delta^{ab}
\end{align}
where $h(g)=30$, 
$f^{abc}$ with $a=1,2,\cdots,496$ 
are the structure constants of 
$SO(32)$ and $\psi^2$ denotes the length of 
the root which is normalized as two. 

The normalization of the Abelian gauge groups 
are estimated by the current algebra or Ka\v{c}-Moody 
algebra of $SO(32)$ which is given by 
\begin{align}
[j_m^a, j_n^b]=i\sum_c f^{abc}j_{m+n}^c 
+\frac{2k}{\psi^2}m \delta^{ab}\delta_{m,-n},
\end{align}
where $k$ is the level of Ka\v{c}-Moody algebra and 
$j_m^a$ are the Laurent coefficients of 
the current $j^a(z)$,
\begin{align}
j^a(z)=\frac{1}{2}N(\psi^i T_{ij}^a\psi^j) 
=\sum_{m=-\infty}^{\infty}\frac{j_m^a}{z^{m+1}},
\end{align}
with $\psi^i$ and $(T^a)_{ij}$ ($i=1,2,\cdots,32)$ being 
the $32$ real fermions and generators in the vector 
representation of $SO(32)$, respectively. 
$N(\psi^i T_{ij}^a\psi^j)$ stands for the normal ordering of the 
operator, $(\psi^i T_{ij}^a\psi^j) $.  
When the level of Ka\v{c}-Moody algebra is equal to one, 
we obtain the operator product expansion of the current
\begin{align}
j^a(z)j^b(w) \sim \frac{2\delta^{ab}}{\psi^2 (z-w)^2} 
+\frac{if^{abc}}{z}j^c(w),
\end{align}
and then we can extract the normalization of 
$(T^a)_{ij}$ as ${\rm tr}(T^aT^b)=2\delta_{ab}$. 

In our model, the generators of $U(1)_{a}$, $T_{a}$ are normalized as 
\begin{align}
&T_1=\frac{1}{\sqrt{2}}{\rm diag}(0,0,0,0,1,1,0,0,\cdots, 0),
\nonumber\\
&T_2=\frac{1}{2}{\rm diag}(1,1,1,1,0,0,0,0,\cdots, 0),
\nonumber\\
&T_3=\frac{1}{\sqrt{12}}{\rm diag}(1,1,1,-3,0,0,0,0,\cdots, 0),
\nonumber\\
&T_4={\rm diag}(0,0,0,0,0,0,1,0,\cdots, 0),
\nonumber\\
&T_5={\rm diag}(0,0,0,0,0,0,0,1,0,\cdots, 0),
\nonumber\\
&\qquad \vdots
\nonumber\\
&T_{13}={\rm diag}(0,0,0,0,0,0,0,0,\cdots, 1),
\end{align}
on the basis of $U(16)$ which is the 
maximal subgroup of $SO(32)$. In general, the  
generators of $U(N)$ can be identified as the part of $SO(2N)$ 
generators. (See e.g. Ref.~\cite{g.ross}.)

\section{The trace identities}
\label{app:trace}

Here, we summarize the trace identities
\begin{align}
&{\rm Tr}F^2 
= 30{\rm tr}F^2 =60F_{SU(3)_C}^2 +
60F_{SU(2)_L}^2 +60\sum_{a=1}^{13} f_a^2,
\nonumber\\
&{\rm Tr}{\bar F}^2 
= 30{\rm tr}{\bar F}^2 =60\sum_{a=1}^{13}{\bar f}_a^2,
\nonumber\\
&{\rm Tr}F{\bar F} 
= 30{\rm tr}F{\bar F} =60\sum_{a=1}^{13}f_a{\bar f}_a,
\nonumber\\
&{\rm tr}F^2{\bar F}^2 
=\left(\frac{1}{2}{\rm tr}(T_2^2)\bar{f}_2^2 +
\sqrt{\frac{{\rm tr}(T_2^2){\rm tr}(T_3^2)}{3}}{\bar f}_2{\bar f}_3 +
\frac{1}{6}{\rm tr}(T_3^2)\bar{f}_3^2
\right) {\rm tr}(F_{SU(3)}^2)
+{\rm tr}(T_1^2){\bar f}_1^2{\rm tr}(F_{SU(2)}^2)
\nonumber\\
&\hspace{1.5cm}+2{\rm tr}(T_1^4){\bar f}_1^2f_1^2 
+2\sum_{c=4}^{13}{\rm tr}(T_c^4)f_c^2\bar{f}_c^2
\nonumber\\
&\hspace{1.5cm}+
2\left( {\rm tr}(T_2^4)\bar{f}_2^2 
+{\rm tr}(T_2^2T_3^2)\bar{f}_3^2 \right) f_2^2
+
4\left( 2{\rm tr}(T_2^2T_3^2)\bar{f}_2\bar{f}_3 
+{\rm tr}(T_2T_3^3)\bar{f}_3^2 \right) f_2f_3
\nonumber\\
&\hspace{1.5cm}+
2\left( {\rm tr}(T_3^4)\bar{f}_3^2 
+{\rm tr}(T_2^2T_3^2)\bar{f}_2^2 
+2{\rm tr}(T_2T_3^3)\bar{f}_2\bar{f}_3 
\right) f_3^2,
\nonumber\\
&{\rm tr}F{\bar F}^3 =
2{\rm tr}T_1^4{\bar f}_1^3f_1 +
2\left({\rm tr}T_2^4{\bar f}_2^3 +
3({\rm tr}T_2^2T_3^2){\bar f}_2{\bar f}_3^2 +
({\rm tr}T_2T_3^3){\bar f}_3^3 
\right) f_2 
\nonumber\\
&\hspace{1.5cm}+\left(
{\rm tr}T_3^4{\bar f}_3^3 +
3({\rm tr}T_2T_3^3){\bar f}_2{\bar f}_3^2 +
3({\rm tr}T_2^2T_3^2){\bar f}_2^2{\bar f}_3
\right) f_3 
+2\sum_{c=4}^{13}{\rm tr}T_c^4{\bar f}_c^3f_c,
\end{align}
where $f_a$ and ${\bar f}_a$ denote the four-dimensional 
and extra-dimensional field strengths of $U(1)_a$ and 
we employ the trace identities such as 
\begin{align}
&{\rm Tr}F^2=30\,{\rm tr}F^2,
\nonumber\\
&{\rm Tr}F^4 = 24{\rm tr}F^4 +3({\rm tr}F^2)^2,
\nonumber\\ 
&{\rm Tr}F{\bar F}^3 = 24{\rm tr}F{\bar F}^3 
+3({\rm tr}F{\bar F})({\rm tr}{\bar F}^2),
\nonumber\\
&{\rm Tr}F^2{\bar F}^2 = 24{\rm tr}F^2{\bar F}^2 
+2({\rm tr}F{\bar F})^2+({\rm tr}F^2)({\rm tr}{\bar F}^2),
\nonumber\\
&{\rm tr}T_1^4=1/2,\,\,{\rm tr}T_2^4=1/4,\,\,{\rm tr}T_3^4=7/12,\,\,
{\rm tr}T_a^4=1\,\,(c=4,\cdots, 13),
\nonumber\\
&{\rm tr}T_2^2T_3^2=1/4,\,\,{\rm tr}T_2^3T_3=0,\,\,
{\rm tr}T_2T_3^3=-1/2\sqrt{3}.
\end{align}


\begin{thebibliography}{99}
\bibitem{Witten:1984dg}
  E.~Witten,
  Phys.\ Lett.\ B {\bf 149} (1984) 351.

\bibitem{Ibanez:2012zz} 
  L.~E.~Ibanez and A.~M.~Uranga,
  Cambridge, UK: Univ. Pr. (2012) 673 p

  
\bibitem{Nilles:2006np}
  H.~P.~Nilles, S.~Ramos-Sanchez, P.~K.~S.~Vaudrevange and A.~Wingerter,
  JHEP {\bf 0604} (2006) 050
  [hep-th/0603086],
  S.~Ramos-Sanchez,
  Fortsch.\ Phys.\  {\bf 10} (2009) 907
  [arXiv:0812.3560 [hep-th]].
  
\bibitem{Nibbelink:2012de}
  S.~Groot Nibbelink and P.~K.~S.~Vaudrevange,
  JHEP {\bf 1303} (2013) 142
  [arXiv:1212.4033 [hep-th]].
  
\bibitem{Friedman:1997ih}
  R.~Friedman, J.~W.~Morgan and E.~Witten,
  alg-geom/9709029,

  R.~Friedman, J.~Morgan and E.~Witten,
  Commun.\ Math.\ Phys.\  {\bf 187} (1997) 679
  [hep-th/9701162].

\bibitem{Blumenhagen:2005zg}
  R.~Blumenhagen, G.~Honecker and T.~Weigand,
  JHEP {\bf 0510} (2005) 086
  [hep-th/0510049].

\bibitem{Donagi:2000zf}
  R.~Donagi, B.~A.~Ovrut, T.~Pantev and D.~Waldram,
  JHEP {\bf 0108} (2001) 053
  [hep-th/0008008],

  B.~Andreas, G.~Curio and A.~Klemm,
  Int.\ J.\ Mod.\ Phys.\ A {\bf 19} (2004) 1987
  [hep-th/9903052].

  V.~Bouchard and R.~Donagi,
  Phys.\ Lett.\ B {\bf 633} (2006) 783
  [hep-th/0512149].

  V.~Braun, Y.~H.~He, B.~A.~Ovrut and T.~Pantev,
  Phys.\ Lett.\ B {\bf 618} (2005) 252
  [hep-th/0501070].
  
\bibitem{Choi:2009pv} 
  K.~S.~Choi, T.~Kobayashi, R.~Maruyama, M.~Murata, Y.~Nakai, H.~Ohki and M.~Sakai,
  Eur.\ Phys.\ J.\ C {\bf 67}, 273 (2010)
  [arXiv:0908.0395 [hep-ph]];
  %
%
  T.~Kobayashi, R.~Maruyama, M.~Murata, H.~Ohki and M.~Sakai,
  JHEP {\bf 1005}, 050 (2010)
  [arXiv:1002.2828 [hep-ph]].
  
\bibitem{Ibanez:1986xy}
  L.~E.~Ibanez and H.~P.~Nilles,
  Phys.\ Lett.\ B {\bf 169} (1986) 354.
  
  
\bibitem{Polchinsky}
  J.~Polchinski, String theory. Vol. 2: Superstring theory and beyond. Cambridge University Press, Cambridge,UK, (1998).   
  
\bibitem{Blumenhagen:2005ga}
  R.~Blumenhagen, G.~Honecker and T.~Weigand,
  JHEP {\bf 0506} (2005) 020
  [hep-th/0504232], 
  
  R.~Blumenhagen, G.~Honecker and T.~Weigand,
  JHEP {\bf 0508} (2005) 009
  [hep-th/0507041].  
  
\bibitem{Weigand:2006yj}
  T.~Weigand,
  Fortsch.\ Phys.\  {\bf 54} (2006) 963.
  
\bibitem{Witten:1995gx}
  E.~Witten,
  Nucl.\ Phys.\ B {\bf 460} (1996) 541
  [hep-th/9511030].
  
\bibitem{Duff:1996rs}
  M.~J.~Duff, R.~Minasian and E.~Witten,
  Nucl.\ Phys.\ B {\bf 465} (1996) 413
  [hep-th/9601036].
  
\bibitem{Aldazabal:1997wi}
  G.~Aldazabal, A.~Font, L.~E.~Ibanez, A.~M.~Uranga and G.~Violero,
  Nucl.\ Phys.\ B {\bf 519} (1998) 239
  [hep-th/9706158].
      
\bibitem{Witten:1982fp}
  E.~Witten,
  Phys.\ Lett.\ B {\bf 117} (1982) 324.

\bibitem{Witten:1985xe}
  E.~Witten,
  Commun.\ Math.\ Phys.\  {\bf 100} (1985) 197.    
      
\bibitem{Blumenhagen:2006ux}
  R.~Blumenhagen, S.~Moster and T.~Weigand,
  Nucl.\ Phys.\ B {\bf 751} (2006) 186
  [hep-th/0603015].    
      
\bibitem{Cremades:2004wa}
  D.~Cremades, L.~E.~Ibanez and F.~Marchesano,
  JHEP {\bf 0405} (2004) 079
  [hep-th/0404229].

  
\bibitem{Fischler:1981zk}
  W.~Fischler, H.~P.~Nilles, J.~Polchinski, S.~Raby and L.~Susskind,
  Phys.\ Rev.\ Lett.\  {\bf 47} (1981) 757.
 
\bibitem{Font:1989aj}
  A.~Font, L.~E.~Ibanez, F.~Quevedo and A.~Sierra,
  Nucl.\ Phys.\ B {\bf 331} (1990) 421.
 
\bibitem{Blumenhagen}
R. Blumenhagen and E. Plauschinn, Introduction to Conformal 
Field Theory: With Applications to String theory, Lect. Notes Phys. 
779 (Springer, Berlin Heidelberg, 2009), 265 p.

\bibitem{g.ross}
Graham G. Ross, Grand Unified Theories, Frontiers 
in physics, 060 (Benjamin/Cummings Publication 
Company, 1984), 235-239 p. 

  
\bibitem{Witten:1998cd}
  E.~Witten,
  JHEP {\bf 9812} (1998) 019
  [hep-th/9810188].

\bibitem{Uranga:2000xp}
  A.~M.~Uranga,
  Nucl.\ Phys.\ B {\bf 598} (2001) 225
  [hep-th/0011048].
  
\bibitem{Garcia-Etxebarria:2014txa}
  I.~Garcia-Etxebarria, M.~Montero and A.~Uranga,
  Phys.\ Rev.\ D {\bf 90} (2014) 12,  126002
  [arXiv:1405.0009 [hep-th]].

\bibitem{Hellerman:2004zm}
  S.~Hellerman,
  hep-th/0405041.
  
\bibitem{Abe:2008fi} 
  H.~Abe, T.~Kobayashi and H.~Ohki,
  JHEP {\bf 0809}, 043 (2008)
  [arXiv:0806.4748 [hep-th]].
  
\bibitem{Abe:2013bca} 
  T.~H.~Abe, Y.~Fujimoto, T.~Kobayashi, T.~Miura, K.~Nishiwaki and M.~Sakamoto,
  JHEP {\bf 1401}, 065 (2014)
  [arXiv:1309.4925 [hep-th]];
%
  Nucl.\ Phys.\ B {\bf 890}, 442 (2014)
  [arXiv:1409.5421 [hep-th]].
  

\end{thebibliography}
\end{document}